\documentclass[11pt,a4paper]{article}
\usepackage[truedimen,margin=28mm]{geometry} 

\usepackage{authblk}

\usepackage{titlesec}
\titleformat*{\section}{\large\bfseries}
\titleformat*{\subsection}{\it}

\usepackage{mathrsfs}
\usepackage{booktabs}
\usepackage{amssymb}
\usepackage{amsmath}
\usepackage{mathtools}
\usepackage{ascmac}
\usepackage{amsthm}
\usepackage[pdftex]{graphicx}
\usepackage[pdftex]{color}
\usepackage{natbib}
\usepackage{setspace}
\usepackage{bm}
\usepackage{url}
\usepackage{color}
\usepackage{tabularx}
\usepackage{ulem}
\usepackage{hyperref}

\usepackage{algorithm}
\usepackage{algorithmic}

\newtheorem{proposition}{Proposition}

\theoremstyle{definition}

\newcommand{\mG}{\mathcal{G}}

\newcommand{\mP}{\mathcal{P}}

\newcommand{\mX}{\mathcal{X}}

\newcommand{\GP}{\mG\mP}

\usepackage{tikz}

\title{{\bf Similarity-Based Random Partition Distribution for Clustering Functional Data}}

\author[1]{Tomoya Wakayama\thanks{Corresponding Author: tom-w9@g.ecc.u-tokyo.ac.jp} }
\author[2]{Shonosuke Sugasawa}
\author[3]{Genya Kobayashi}

\affil[1]{Graduate School of Economics, The University of Tokyo}
\affil[2]{Faculty of Economics, Keio University}
\affil[3]{School of Commerce, Meiji University}
\date{}

\begin{document}
\maketitle

\begin{abstract}

Random partition distribution is a crucial tool for model-based clustering.
This study advances the field of random partition in the context of functional spatial data, focusing on the challenges posed by hourly population data across various regions and dates. We propose an extension of the generalized Dirichlet process, named the similarity-based generalized Dirichlet process (SGDP)-type distribution, to address the limitations of simple random partition distributions (e.g., those induced by the Dirichlet process), such as an overabundance of clusters. This model prevents excess cluster production and incorporates pairwise similarity information to ensure accurate and meaningful clustering. The theoretical properties of the SGDP-type distribution are studied. Then, SGDP-type random partition is applied to a real-world dataset of hourly population flow in $500\text{m}$ meshes in the central part of Tokyo. In this empirical context, our method excels at detecting meaningful patterns in the data while accounting for spatial nuances. The results underscore the adaptability and utility of the method, showcasing its prowess in revealing intricate spatiotemporal dynamics. The proposed random partition will significantly contribute to urban planning, transportation, and policy-making and will be a helpful tool for understanding population dynamics and their implications.

\end{abstract}

\noindent%

\section{Introduction}\label{sec:intro}
The past few decades have witnessed a rapid proliferation of mobile devices. The resulting surge in fine population data is pivotal for comprehending and planning the foundational aspects of contemporary society.
Specifically, it encompasses various domains, including urban and transportation planning, healthcare service distribution, and extensive policy-making~\citep{paez2004spatial,wang2018spatial,ahmadi2018location}. The collection and analysis of population data also hold economic significance. By meticulously analysing population data, understanding consumer behavioural patterns and preferences becomes feasible, which aids in developing and executing marketing strategies~\citep{pol1986marketing,nagata2013using}. Thus, effectively utilizing population statistics is imperative for societal advancement.

\begin{figure}[t]
  \centering
  \includegraphics[width=0.6\linewidth]{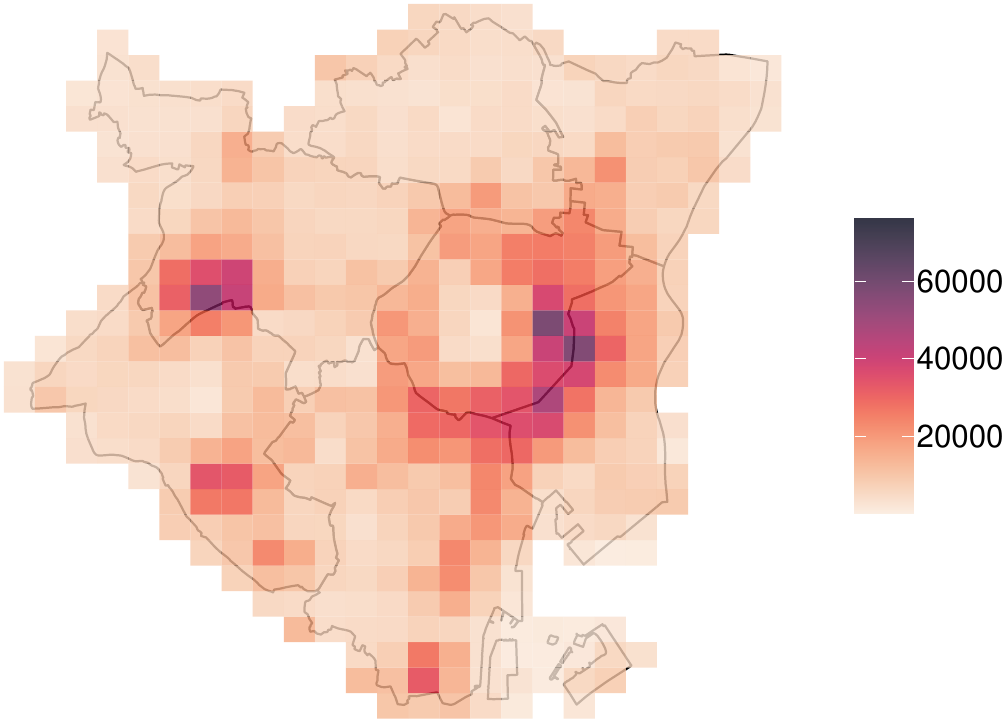}
  \caption{Population in the central districts of Tokyo at 2 PM on January $29, 2019$ (Japan Standard Time, JST). \label{fig:districts}}
\end{figure}

Clustering is a viable technique to distill meaningful insights from population data. Identifying clusters and commonalities within clusters and unveiling regional traits can contribute to both industry, such as ridesharing services and street advertising, and research, including urban engineering and humanities.
In particular, model-based clustering is compatible with nonparametric Bayesian methods. Dirichlet process (DP)-based clustering, as introduced by~\cite{ferguson1973bayesian,ferguson1974prior}, can autonomously ascertain cluster~quantities and incorporate spatial structures. The seminal study of \cite{dahl2017random} adeptly integrated geographical proximity when assigning existing clusters to new items within the Pitman--Yor process~\citep{ishwaran2001gibbs,pitman1997two}, contributing to valuable applications~\citep{glynn2021inflection,grazian2023review}. Notably, adjacency information is crucial in population data analysis~\citep{lym2021exploring,zhang2019prefecture}, and a high correlation is observed between neighbouring districts, as depicted in Figure \ref{fig:districts}.

\begin{figure}[t]
    \centering
    \includegraphics[width=0.95\linewidth]{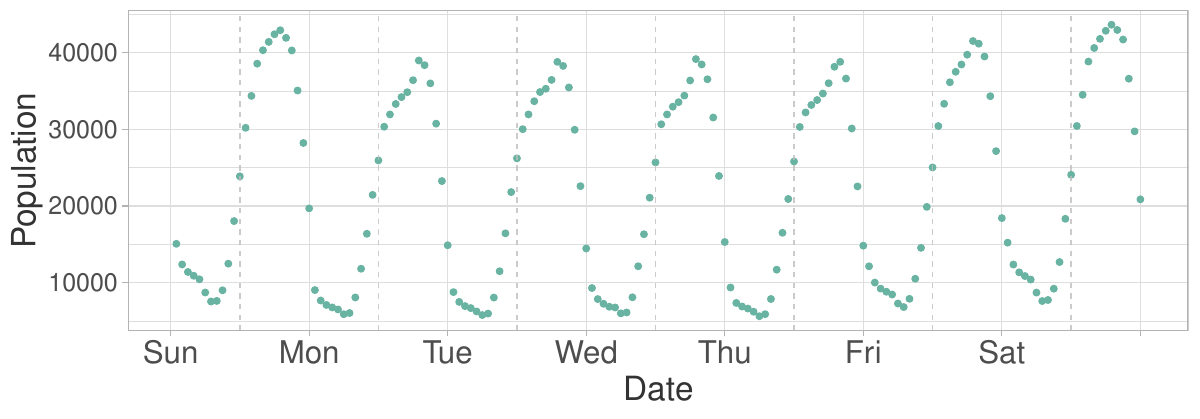}
    \caption{Hourly population flow in a week in a certain mesh.\label{fig:pop_flow}}
\end{figure}

Nonetheless, two primary attributes of our dataset--its multivariate nature and the presence of temporal information--impede its adoption by existing methods. When tackling multivariate clustering problems, applying methods that do not fix the number of clusters requires ingenuity; otherwise, it leads to the creation of excess clusters and is inappropriate. Owing to their inherent complexity, high-dimensional data are readily classifiable, a trait that aids in supervised classification tasks~\citep{delaigle2012achieving,wakayama2021fast}. Conversely, this high classifiability may lead to an excessive number of clusters in unsupervised clustering. In particular, this property would accelerate the tendency of the DP to favor smaller cluster sizes and create superfluous clusters~\citep{miller2013simple,miller2014inconsistency}, thereby hindering the extraction of beneficial knowledge from the data. Shifting the focus to the temporal aspect reveals that considering the spatial structures alone is inadequate. Figure \ref{fig:pop_flow} depicts the hourly population flow of a week in a specific area. The intra-day pattern may vary between weekdays and holidays or owing to pre-holiday effects such as the ``Happy Friday'' phenomenon~\citep{stutz2004charting, lu2012strategic}. In such scenarios, cluster structures vary over time, and vital insights could be missed by focusing only on typical weekday trends.

To address these challenges, we propose a similarity-based random partition distribution using a generalized Dirichlet process~\citep[GDP,][]{hjort2000bayesian}, namely similarity-based GDP (SGDP) type random partition, and cluster the observed hourly population flow as functional data, which are realizations of stochastic processes. Our methodology reduces the risk of over-clustering through generalized parameterization \citep{rodriguez2014functional}, and it integrates geographic adjacency, which is explored theoretically in this study. Furthermore, we model the discrepancy between observations and the cluster mean using a Gaussian process. Intuitively, if a value deviates from the cluster mean at a particular time in a region, then the effect naturally spreads to the contiguous times as well. Technically, if the gap is assumed to be independent noise, then the similarity (likelihood) between the observation and the cluster mean will be overly small, resulting from the calculated product of the likelihoods for the number of observations. However, this issue is less pronounced in Gaussian processes, as they account for correlations. Additionally, our model adapts to temporal variations, enabling the identification of time-series cluster shifts. This approach facilitates a nuanced capture of the distinctive characteristics of each region.

The proposed method differs from existing model-based nonparametric Bayesian clustering in some key aspects. Several approaches focus on univariate observations and employ parametric regression structures that are subsequently clustered using random partition for random effects or other parameters \citep[e.g.,][]{mozdzen2022bayesian, cremaschi2023change}. Our methodology embraces a fully nonparametric model and has constructed the clustering distribution to incorporate spatial structure and control the number of clusters.

The remainder of this paper is structured as follows. Section~\ref{sec:method} reviews the GDP, delineates a similarity-based random partition and discusses its properties. In Section~\ref{sec:SGDP}, we introduce the proposed model for clustering spatial functional data, detailing the prior distribution setup and the computation of the posterior distribution. Sections~\ref{sec:sim} and \ref{sec:app} describe the simulation experiments and apply the model to population data, presenting the empirical findings. Section~\ref{sec:dis} discusses the major conclusions drawn from the study findings and presents future research directions. The Julia code used to implement the proposed methods is publicly available in the GitHub repository (https://github.com/TomWaka/Similarity-based-Generalized-Dirichlet-Process).

\section{Similarity-based random partition distribution}\label{sec:method}

\subsection{GDP and induced partition distribution}\label{sec:review}

To construct the GDP, we employ a stick-breaking construction, reflective of its discrete nature \citep{hjort2000bayesian,rodriguez2014functional}. A probability measure $G$ is defined as a GDP if it follows the following formulation:
\begin{align*}
    G(\cdot) = \sum_{h=1}^{\infty} w_h \delta_{m_h}(\cdot) ,
\end{align*}
where $w_h= v_h\prod_{\ell=1}^{h-1}(1-v_{\ell})$, $v_h\sim Be(\alpha\beta,\alpha(1-\beta))$, $m_h\sim G_0$, $\alpha\in \mathbb{R}_+$, and $\beta \in (0,1)$.
$G_0$ refers to a nonatomic probability measure, often termed a base measure of $G$, and all $\{v_h\}_{h\ge 1}, \{m_h\}_{h\ge 1}$ are independent. In this configuration, we denote the probability distribution as $GDP(\alpha\beta,\alpha(1-\beta), G_0).$ In the scenario where $\alpha\beta=1$, $G$ is reduced to the DP~\citep[][]{ferguson1973bayesian,ferguson1974prior,sethuraman1994constructive,ishwaran2001gibbs,Muller2015}. 

Suppose we have $n$ items $\{1,2,\ldots,n\}$ independently sampled from $GDP(\alpha\beta,\alpha(1-\beta), G_0)$ and partitioned into $k$ distinct clusters $(N_1,\ldots,N_k)$, where $N_j$ represents the cardinality of the $j$th cluster for $j=1,\ldots,k$.
The predictive probability function of assignment $z_{n+1}$ can be explicitly obtained as follows \citep{pitman1995exchangeable, barcella2018dependent}:
\begin{equation}\label{eqn:GDP-ch}
\begin{aligned} 
&p(z_{n+1}=j \mid z_1,\ldots,z_n)  = 
    \frac{\alpha\beta + N_j - 1}{\alpha + n - 1} 
    \prod_{\ell=1}^{j-1} 
    \left( 
        \frac{\alpha(1 - \beta) + \sum_{m=\ell+1}^k N_m}{\alpha + \sum_{m=\ell+1}^k N_m - 1} 
    \right),
    \  (1 \le j \le k), \\
&p(z_{n+1}=k+1 \mid z_1,\ldots,z_n)  = 
    \frac{\alpha(1 - \beta)}{\alpha + n - 1} 
    \prod_{\ell=1}^{k-1} 
    \left( 
        \frac{\alpha(1 - \beta) + \sum_{m=\ell+1}^k N_m}{\alpha + \sum_{m=\ell+1}^k N_m - 1} 
    \right).
\end{aligned}
\end{equation}
where $z_1,\ldots,z_n$ are the clustering assignments, $N_j=\sum_{i=1}^n I(z_i=j)$, and $I(\cdots)$ is an indicator function. The expression \eqref{eqn:GDP} comprises two scenarios: the first detailing the probability of a new item being stored as an existing cluster, and the second regarding the probability of it being assigned to a new cluster. Note that \eqref{eqn:GDP-ch} is not exchangeable and order-dependent: the probability for the $(n+1)$-th observation changes if the conditioning data are permuted. This issue will be addressed in Section~\ref{sec:SGDP}. Expression~\eqref{eqn:GDP-ch} demonstrates that the joint probability of $(z_1,\ldots,z_n)$ can be depicted as 
$$
p(z_1,\ldots,z_n; \alpha, \beta)=\prod_{i=1}^n p(z_i\mid z_1,\ldots,z_{i-1}; \alpha, \beta), 
$$
where 
\begin{equation}\label{eqn:GDP}
\begin{aligned} 
&p(z_i=j\mid z_1,\ldots,z_{i-1};\alpha,\beta)  = \frac{\alpha\beta+N_j(i)-1}{\alpha+i-2} \prod_{\ell=1}^{j-1} A_\ell(i),
    \  (1 \le j \le k), \\
&p(z_i=k+1\mid z_1,\ldots,z_{i-1};\alpha, \beta)  = 
    \frac{\alpha(1-\beta)}{\alpha+i-2} \prod_{\ell=1}^{k-1} A_\ell(i),
\end{aligned}
\end{equation}
and 
$$
A_\ell(i)= \frac{\alpha-\alpha\beta+\sum_{m=\ell+1}^k N_m(i) }{\alpha-1+\sum_{m=\ell+1}^k N_m(i) }, \ \ \ell=1,\ldots,j-1,
$$
where $k=k(i)$ denotes the number of clusters induced by $z_1,\ldots,z_{i-1}$, and $N_m(i)$ indicates the size of the $m$th cluster induced by $z_1,\ldots,z_{i-1}$. When $\alpha\beta=1$, $A_\ell(i)=1$ for all $\ell$ and $i$, suggesting that the distribution~\eqref{eqn:GDP-ch} is equal to the Ewens distribution~\citep{ewens1972sampling,pitman1995exchangeable,pitman1996some}.

The number of clusters in the partitions constructed by the GDP depends on $n$, $\alpha$, and $\beta$. \cite{rodriguez2014functional} proved that when $\alpha\beta>1$, the expected number of clusters (say, $E[K_n]$) remains finite even as the number of observations $n$ diverges. Intuitively, because the second case of \eqref{eqn:GDP} is a decreasing function of $\alpha\beta$, if $\alpha\beta$ is large, a new cluster is unlikely to be generated, indicating a significant departure from the scenario $\alpha\beta=1$, that is, the standard DP, where $k\approx \log n$ as $n$ approaches infinity~\citep{korwar1972contributions,antoniak1974mixtures}.
Such flexibility in controlling the growth in the number of clusters is crucial for limiting an excessive number of clusters during clustering.

\subsection{Introducing pairwise similarity in random partition}
Next, we extend the GDP-type random partition to incorporate pairwise information. Here, we consider the scenario where pairwise similarity information $s_{ii'}(\tau)$, such as covariate distance or contingency information, exists for each pair of items $i, i' = 1,\ldots,n$. In the application described in Section~\ref{sec:app}, we define $s_{ii'}=1$ for adjacent areas and $s_{ii'}=\tau \in(0,1)$ otherwise. The objective is to embed $s_{ii'}$ into the prior distribution of $z_1,\ldots,z_n$ such that two subjects with large values of $s_{ii'}$ are more likely to belong to the same cluster. We then extend the conditional probability given in \eqref{eqn:GDP-ch} as the following SGDP-type random partition distribution:
\begin{equation}\label{SDGP}
p_\omega(z_i=j\mid z_1,\ldots,z_{i-1}; \alpha,\beta)
=\omega_j(i)\frac{\alpha\beta+N_j(i)-1}{\alpha+i-2} \prod_{\ell=1}^{j-1} A_\ell(i),
\end{equation}
for the $i$th item as an existing cluster $j\in \{1,\ldots,k\}$, where 
\begin{equation}\label{eq:omega-omega*}
\omega_j(i)=\left(\frac{\sum_{j'=1}^k (\alpha\beta+N_{j'}(i)-1)\prod_{\ell=1}^{j'-1}A_\ell(i)}{\sum_{j'=1}^k \omega_{j'}^{\ast}(i)(\alpha\beta+N_{j'}(i)-1)\prod_{\ell=1}^{j'-1}A_\ell(i)}\right)\omega_j^{\ast}(i),
\end{equation}
and 
\begin{equation}\label{eq:similarity}
    \omega_j^{\ast}(i)=\frac{\sum_{i'=1}^{i-1}I(z_{i'}=j)\lambda(s_{ii'})}{\sum_{i'=1}^{i-1}\lambda(s_{ii'})},
\end{equation}
where $\lambda(\cdot):\mathbb{R}\to\mathbb{R}$ denotes an increasing function. Importantly, the transformation \eqref{eq:omega-omega*} of the original similarity weight $\omega_j^{\ast}(i)$ to $\omega_j(i)$ lets \eqref{SDGP} act as a proper probability distribution and does not impact the relative magnitude. Indeed, $\omega_j(i)$ equals $\omega_j^{\ast}(i)$ multiplied by some constant factor independent of $j$ ($\omega_j(i)\propto \omega_j^{\ast}(i)$ for any $i$).

$\lambda$ can be any increasing function, and we adopt the identity function in our experiments in Sections~\ref{sec:sim}~and~\ref{sec:app}. If $s_{ii'}$ changes continuously according to the distance between regions, it would be appropriate to consider other functions, such as exponential or polynomial functions.

\subsection{Properties of SGDP}\label{subsec:prop}
Here, we examine two key properties of the SGDP-type random partition distribution as defined in equation \eqref{SDGP}: first, the probability of generating a novel cluster, and second, the role that the similarity measure plays in the allocation process.

First, we focus on the probability of creating a new cluster. For the proposed distribution, the following expression holds from equation~\eqref{eqn:GDP}: 
\begin{align*}
    \sum_{j=1}^k p_\omega(z_i=j\mid z_1,\ldots,z_{i-1}; \alpha,\beta) 
     =\sum_{j=1}^k \frac{\alpha\beta+N_j(i)-1}{\alpha+i-2} \prod_{\ell=1}^{j-1} A_\ell(i),
\end{align*}
for $i=1,\ldots,n$, aligning with the probability of assigning $z_i$ to existing clusters under the standard GDP (absence of the similarity measure). Hence, the conditional probability of assigning $z_n$ to a new cluster is expressed in \eqref{eqn:GDP-ch}, which remains unaffected by the similarity weight $\omega_j^{\ast}(i)$. Consequently, the similarity measure does not affect the growth in the number of clusters, and the role of the similarity measure is independent of those of $\alpha$ and $\beta$; this feature is preferable in terms of the interpretability of the parameters \citep{dahl2017random}.
Specifically, for a given cluster, the probability of a new cluster occurring for GDP- and SGDP-type partition is
\begin{align*}
     p_\omega(z_{i}=k+1\mid z_1,\ldots,z_{i-1}; \alpha,\beta) 
     = 
      \frac{\alpha-\alpha\beta}{\alpha+i-2} \prod_{\ell=1}^{k-1} \frac{\alpha-\alpha\beta+\sum_{m=\ell+1}^k N_m }{\alpha- 1 +\sum_{m=\ell+1}^k N_m },
\end{align*}
and it is indeed irrespective of $\lambda$ and $\tau$. If $\beta\in(0,1)$ is large, then the first term becomes small. Additionally, when $\alpha\beta>1$, each term of the $k-1$ products is less than $1$, indicating that $\alpha$ and $\beta$ are pivotal in controlling the cluster quantity during SGDP-type partitioning.

Subsequently, we assess the impact of the similarity measure (adjacency structure). According to the formulation of \eqref{SDGP}, similarity does not affect the probability of new cluster creation for each allocation. We designed \eqref{SDGP} to emphasize spatial proximity when assigning the cluster to which a new item should be assigned among existing clusters. This feature is corroborated by the following two results.
\begin{proposition}\label{pro1} 
For fixed parameters $\alpha>0,\beta\in(0,1),\tau>0$, and any partition of $(z_1,\ldots,z_{i-1})$, the prior probability \eqref{SDGP} of a new item being assigned to a particular cluster, $p_\omega(z_i=j\mid z_1,\ldots,z_{i-1}; \alpha,\beta,\tau)$, is an increasing function of the number of items in that cluster adjacent to the new item.
\end{proposition}

\begin{proof}
To simplify the notation, Equation~\eqref{SDGP} is expressed as follows:
\begin{equation*}
    p_\omega(z_i=j\mid z_1,\ldots,z_{i-1}; \alpha,\beta,\tau) = C_{1,ij}\ \omega_j(i).
\end{equation*}
In particular, $C_{1,ij}>0$ does not include similarity information. Thereafter, we obtain
\begin{align*}
    \omega_j(i) & = \left(\frac{\sum_{j'=1}^k (\alpha\beta+i-2)\prod_{\ell=1}^{j'-1}A_\ell(i)}{\sum_{j'=1}^k \omega_{j'}^{\ast}(i)(\alpha\beta+i-2)\prod_{\ell=1}^{j'-1}A_\ell(i)}\right)\omega_j^{\ast}(i)\\
    & = C_{2,i}\  \omega_j^{\ast}(i),
\end{align*}
where $C_{2,i}>0$ is constant over $j=1,2,\ldots,k$. Hence, the role of similarity information in allocation is solely through $\omega_j^{\ast}(i)$, attributed to the following definition: 
\begin{equation*}
    \omega_j^{\ast}(i)=\frac{\sum_{i'=1}^{i-1}I(z_{i'}=j)\lambda(s_{ii'})}{\sum_{i'=1}^{i-1}I(z_{i'}=j)\lambda(s_{ii'})+\sum_{i'=1}^{i-1}I(z_{i'}\neq j)\lambda(s_{ii'})}.
\end{equation*}
$p_\omega(z_i=j\mid z_1,\ldots,z_{i-1}; \alpha,\beta,\tau)$ increases with the number of items adjacent to item $i$ within cluster $j$.
\end{proof}

\begin{proposition}
Assume that $z_1,\dots,z_{i-1}$ are distributed to \eqref{SDGP}, and let the hyperparameters $\alpha, \beta, \text{ and } \tau$ be random variables with specified prior distributions. Then, the marginal prior probability of a new element $z_i$ being assigned to a particular cluster increases as the number of items in the cluster adjacent to the new item grows.
\end{proposition}

\begin{proof}
Confirming that the derivative of the marginal distribution of $\omega^*_j(i)$ is positive is sufficient. Note that the marginal distribution is the integral of the product of the probability that item $i$ belongs to cluster $j$ given the parameters and the probability distribution of these parameters. Since the derivative of the former probability, $\frac{\partial}{\partial \omega^*_j(i)} p_\omega(z_i=j\mid z_1,\ldots,z_{i-1}; \alpha,\beta,\tau)$, is positive (Proposition~\ref{pro1}), the Leibniz integral rule \citep{folland1999real} implies that the integral is also positive.
\end{proof}

These findings confirm that the prior distribution appropriately incorporates spatial information. Additionally, in the posterior distribution, in the process of aligning data and clusters, inherently close districts can be placed in the same cluster; however, the observed data may deviate slightly (owing to noise).

\subsection{Connection to other distributions}\label{sec:priors}
\begin{figure}[t]
    \centering
    \includegraphics[width=0.8\linewidth]{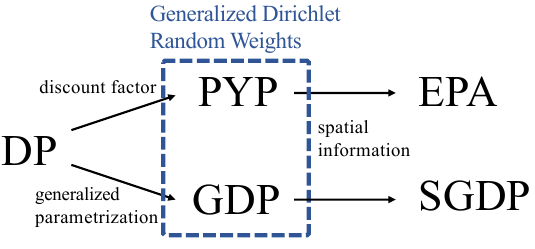}
    \caption{Diagram of random partition distributions. The relationships between Dirichlet process (DP), Pitman--Yor process (PYP), generalized Dirichlet process (GDP), Ewens--Pitman attraction distribution (EPA), and similarity-based GDP type distribution (SGDP).}
    \label{fig:connect}
\end{figure}

We explore the relationship between our partition distribution and existing frameworks, as depicted in Figure~\ref{fig:connect}.
Setting $\alpha\beta=1$ in the proposed distribution yields alignment with the Ewens--Pitman attraction distribution with a discount of $0$, developed by \cite{dahl2017random}.
A fundamental divergence from this approach is our adoption of the GDP rather than the Pitman--Yor process. While GDP and the Pitman--Yor process fall under the same category of Generalized Dirichlet Random Weights~\citep{connor1969concepts,ishwaran2001gibbs}, they are differentiated by their parametrization philosophies. The GDP signifies a straightforward relaxation of the model constraints, whereas the Pitman--Yor process revises the DP's ``rich-get-richer'' paradigm through the incorporation of a discount factor. The choice of model philosophy is predominantly determined by the specific attributes of the dataset and analytical objectives.

Considering the attributes of alternative distributions and datasets is instructive to comprehend the rationale for employing GDP-type methodologies. Primarily, in the DP, the number of clusters increases at a logarithmic rate of the sample size~\citep{korwar1972contributions,antoniak1974mixtures}. Both practical and theoretical evidence suggest that the DP's tendency to finely differentiate data often results in an excessive number of clusters \citep{miller2013simple}. Moreover, the Pitman--Yor process, characterized by its power-law tail decay, encounters similar challenges \citep{pitman1997two,ayed2019beyond,miller2014inconsistency}. While the Pitman--Yor process can yield a finite number of clusters with specific discount parameters, it requires the number of clusters to be fixed in advance; therefore, unless one assumes that the number of components is known a priori, it is not practical~\citep{deblasi2013gibbs,miller2014inconsistency}.
Subsequently, the dataset of primary interest in this context is functional data, notably high-dimensional. Owing to their intricate structures, functional data are readily classifiable, thus benefitting supervised classification tasks \citep{delaigle2012achieving,wakayama2021fast}. However, in clustering scenarios where the number of clusters is indeterminate, such high dimensionality may engender an overabundance of clusters. Consequently, the intrinsic nature of the DP could be exacerbated by the functional data, and thus, using a GDP-type method is preferable \citep{rodriguez2014functional}. 
To mitigate over-clustering, it is preferable to restrict the parameter space to the region where $\alpha\beta > 1$ or to assign larger probability density there. Therefore, in our implementation, we adopt a prior distribution that places more mass in the region $\alpha\beta > 1$, and we also discuss the resulting behavior of the posterior.


\section{Clustering functional data with the SGDP}\label{sec:SGDP}

\subsection{Model settings}
Let $y_1(x),\ldots,y_n(x)$ represent functional observations for $x\in \mathcal{X}$. We consider the following model-based clustering: 
\begin{equation}\label{mixture}
\begin{split}
&y_i(x)\mid\mu_i(x)\sim \GP(\mu_i(x),C_y),\ \ \ \mu_i(x)= \sum_{j=1}^{K_n} \theta_j(x)I(z_i=j),  \\
&\theta_j(x) \sim \GP(m_{\theta},C_{\theta}), \ \ \  (z_1,\ldots,z_n)\sim {\rm SGDP}(\alpha,\beta,\tau), \\
& i=1,\ldots,n, \quad j=1,\ldots,K_n,
\end{split}
\end{equation}
where $\GP$ denotes Gaussian process, $\mu_i(\cdot)$ is the mean function of $y_i(x)$, $C_y(\cdot)$ is the covariance function of the error term, $K_n$ is the number of clusters, $z_1,\ldots,z_n$ are membership variables, $\theta_j(\cdot)$ is the common mean function within the $j$th cluster, $m_{\theta}(\cdot)$ is the overall mean function, and $C_{\theta}(\cdot)$ is the covariance function. Here, ${\rm SGDP}(\alpha,\beta,\tau)$ denotes the SGDP with parameters $\alpha$, $\beta$, and $\tau$, that is, the membership variables follow a prior random partition distribution~\eqref{SDGP}. A gamma prior distribution is placed on $\alpha$, and beta prior distributions are placed on $\beta$ and $\tau$. Regarding $m_{\theta}$, we impose a $\GP (m_m,C_m)$ prior. If we assume a radial basis function (RBF) \citep[e.g.,][]{rasmussen2006gaussian}, $K(x,x')=\eta^2\exp(-\|x-x'\|^2/\phi^2)$, for covariance matrices $C_y$ and $C_{\theta}$, then the scale parameters $\eta_y$ and $\eta_{\theta}$ have conjugate priors: inverse-gamma distributions. As for the range parameters $\phi_y$ and $\phi_{\theta}$, arbitrary prior distributions reflecting analysts' beliefs can be used. If the data are observed repeatedly $T$ times in each area $i$ (e.g., if the functions are observed daily), $y_i$ can be replaced with $y_{it}$ and $\mu
_i$ with $\mu_{it}$ $(t=1,\cdots ,T)$.

The proposed distribution depends on an ordering of the observations. 
Therefore, to address the uncertainty in the ordering, we suggest assuming a uniform prior over all $n!$ for the permutation of the order $\{1,\dots,n\}$. Within the MCMC sampling scheme (detailed in Section~\ref{sec:pos}), we incorporate a Metropolis--Hastings step~\citep{dunson2020hastings, gelman2013bayesian} to sample from the full conditional posterior distribution over permutations. This approach aims to average over the uncertainty associated with data ordering, similar to \cite{dahl2017random}.

\subsection{Posterior computation} \label{sec:pos}
We present an algorithm for simulating the posterior distribution of the nonparametric Bayesian model in \eqref{mixture}. While this model offers a powerful framework for nonlinear regression and clustering, posterior inference is complicated by the infinite-dimensional nature of the Gaussian process and lack of a straightforward stick-breaking representation. To circumvent these challenges, namely the Gaussian process, we consider a finite-dimensional approximation to the Gaussian process based on a finite set of measurement points $\bm{x} \subset \mX$. Then, the joint posterior distribution is represented as follows:
\begin{align*}
    &\prod_{i=1}^n p(y_{i}(\bm{x})\mid\mu_{i}(\bm{x}), C_y)\times \pi(z_{1},\ldots,z_{n},K_{n}=k \mid \alpha, \beta,\tau)\\
    &\times \prod_{j=1}^{K_n} \pi(\theta_{j}(\bm{x})\mid m(\bm{x}),C_{\theta}(\bm{x}) ) \times \pi(\alpha,\beta,\tau,\eta_{\theta},\phi_{\theta},\eta_y,\phi_y).
\end{align*}
While some nonparametric Bayesian models admit efficient posterior simulation via specialized algorithms~\citep{maceachern1998estimating,ishwaran2000markov,neal2000markov,Muller2015}, the non-exchangeability and complex form of our formulation preclude such approaches. Therefore, we resort to a general-purpose Markov chain Monte Carlo (MCMC) technique, the Gibbs sampler \citep{gelfand1990sampling}, which iteratively samples each parameter from its full conditional distribution given the current values of all the other parameters.

Most parameters can be updated by exploiting conjugacy. In particular, the cluster-specific GP atoms $\theta_j$ can be sampled from multivariate Gaussian full conditionals:
\begin{align*}
       p(\theta_{j}\mid \cdot) &\propto \pi(\theta_{j}(\bm{x})\mid m(\bm{x}),C_{\theta}(\bm{x}) )\prod_{i:z_i=j}  p\left( y_i\mid \theta_{j}(\bm{x}), C_y(\bm{x})\right)\\
       &\propto N(m_{pos},\ C_{pos}),
\end{align*}
where $C_{pos} = \{  C_{\theta}^{-1} + N_jC_y^{-1}(\bm{x}) \}^{-1}$ and $m_{pos}=  C_{pos} \{  {C_{\theta}^{} }^{-1} m_{\theta}^{} (\bm{x}) + \sum_{i:z_i=j}  C_y^{-1}(\bm{x})  y_i \}$.

The cluster assignments $z_i$ can be sampled from a discrete full conditional $p(z_{i}=j\mid \cdot)$ of the form:
\begin{align*}
       \begin{cases}
       p(z_{i}=j\mid \bm{z}_{-i},\alpha,\beta)p(y_{i}(\bm{x}) \mid \theta_{j}(\bm{x}), C_y),\ \  j=1,\ldots,k^{-}, \\
       p(z_{i}=j\mid \bm{z}_{-i},\alpha,\beta)p(y_{i}(\bm{x}) \mid m_{\theta}(\bm{x}), C_y+C_{\theta}),\ \  j=k^{-} +1.
       \end{cases}.
\end{align*}
We suggest randomly shuffling data and accepting or rejecting the new permutation by the Metropolis--Hastings algorithm before the assignments to eliminate the effect of the order of observation. Specifically, a new permutation $\pi'$ is proposed from the current permutation $\pi$ (e.g., by randomly selecting two distinct indices $j,k\in\{1,\ldots,n\}$ and swapping the elements). This proposed permutation is accepted with probability 
\[
\min\left(1,\frac{p(y\mid \pi',\cdots )p(\pi')q(\pi\mid\pi')}{p(y\mid \pi,\cdots )p(\pi')q(\pi\mid\pi')}\right),
\]
where $q$ is a proposal distribution. Given the uniform prior $p(\pi)$ and a symmetric proposal $q(\pi'\mid\pi) =q(\pi\mid\pi')$, this simplifies to the ratio of likelihoods.

If prior distributions for $\eta_y^2$ and $\eta_{\theta}^2$ are set to an $IG(\frac{a_{\eta}}{2},\frac{b_{\eta}}{2})$, which denotes an inverse-gamma distribution with shape parameter $a_{\eta}$ and scale parameter $b_{\eta}$, then the full conditional distribution for $\eta_y$ is 
   \begin{align*}
    IG\Bigg(
    \frac{a_{\eta} + N |\mathcal{X}|}{2}, \frac{b_{\eta} + \sum_{i=1}^n 
        \left( y_{i}(\bm{x}) - \mu_{i}(\bm{x}) \right)^{\top} 
        R^{-1}_y(\phi_y) 
        \left( y_{i}(\bm{x}) - \mu_{i}(\bm{x}) \right)}{2}
    \Bigg),
\end{align*}
where $R_y = \eta^{-2}_yC_y$, and the full conditional distribution for $\eta_{\theta}$ is 
\begin{align*}
    IG\Bigg(
    \frac{a_{\eta} + K_n |\mathcal{X}|}{2}, 
    \frac{b_{\eta} + \sum_{j=1}^{K_n} 
        \left( \theta_{j}(\bm{x}) - m(\bm{x}) \right)^{\top} 
        R^{-1}_2(\phi_{\theta}) 
        \left( \theta_{j}(\bm{x}) - m(\bm{x}) \right)}{2}
    \Bigg),
\end{align*}
where $R_{\theta} = {\eta_{\theta}}^{-2}C_{\theta}$.

Finally, the prior mean $m_{\theta}^{}$ has a conjugate Gaussian full conditional $N(m_{pos},C_{pos})$:
    \begin{align*}
    C_{pos} &= \left(K_{n} {C_{\theta}^{} }^{-1} +C_m^{-1}\right)^{-1}, \\ 
    m_{pos} &= C_{pos}  \left\{ {C_{\theta}^{} }^{-1} \sum_{j=1}^{K_n} \theta_{j} + C_m^{-1} m_m \right \}.
    \end{align*}

For parameters lacking conjugate priors, the Metropolis--Hastings algorithm is employed. The posterior distributions are sampled using proposal distributions and acceptance probabilities by assigning suitable priors that reflect the analyst's beliefs. Specifically, $\tau$ is assigned a Beta($a_{\tau},b_{\tau}$) prior, and its posterior is sampled using a Gaussian proposal distribution $\tau^* \mid \tau \sim N(\tau, 10^{-2})$. $\alpha$ is given a Gamma($a_{\alpha},b_{\alpha}$) prior, and a Gaussian proposal distribution $\alpha^* \mid \alpha \sim N(\alpha, 10^{-1})$ is employed for posterior sampling. Similarly, $\beta$ is assigned a Beta($a_{\beta},b_{\beta}$) prior, and its posterior is sampled using a Gaussian proposal distribution $\beta^* \mid \beta \sim N(\beta, 10^{-1})$. The length-scale parameters $\phi_{y}$ and $\phi_{\theta}$ of the covariance kernels are assigned inverse-gamma priors IG($a_{\phi},b_{\phi}$); their posterior distributions are sampled using Gaussian proposal distributions $\phi_{y}^* \mid \phi_{y} \sim N(\phi_{y}, 10^{-1})$ and $\phi_{\theta}^* \mid \phi_{\theta} \sim N(\phi_{\theta}, 10^{-1})$, respectively.

\section{Simulation}\label{sec:sim}
In this section, we evaluate the clustering performance and mean function estimation accuracy of the proposed method through numerical experiments and discuss its effectiveness.

\subsection{Setting}
First, we introduce the data-generating process along with the model in \eqref{mixture}. The experiment contains data with different means depending on clusters. Specifically, observations across $40$ areas for $15$ days are assumed, with each day comprising $24$ data points. In essence, $15$ curve-like observations are gathered in $40$ districts. We divided $40$ areas into $8$ clusters, each containing $5$ areas. For each cluster, a common mean structure $\theta_j$, consisting of $24$ points, was generated using $\mG\mP(0, 2, 5)$, where $\mG\mP(0, \eta, \phi)$ denotes a zero-mean Gaussian process with RBF kernel $K(x,x')=\eta^2\exp(-|x-x'|^2/\phi^2)$ in Case~1 and the exponential kernel $K(x,x')=\eta^2\exp(-|x-x'|/\phi)$ in Case~2 with $x,x'\in\mathbb{R}$. Additionally, we sample $y_{it}$ by adding noise to $\mu_{it}$, which was also generated in each region using either $\mG\mP(0, 1, 1)$ or $\mG\mP(0, 3/2, 1)$. The former is a high signal-to-noise ratio (SNR) case, whereas the latter is a low SNR case. We posit that all the areas within the same cluster are adjacent. Hence, while each area has unique observations, they share a prominent trend within the cluster. Here, the primary objective is to formulate an approach for accurately classifying the generated data based on their general shapes and adjacency structures.

We generate $50$ different datasets using the aforementioned procedure and analyse each dataset using the model in \eqref{mixture} with three distinct random partitions: the SGDP (our proposal), GDP, and SGDP with $\alpha\beta=1$ (SDP, similarity-based Dirichlet process). The prior distributions for the range parameters, $\phi_y$ and $\phi_{\theta}$, and the scale parameters, $\eta_y$ and $\eta_{\theta}$, are set to IG$(1/2, 1/2)$ distribution. Concerning $m_{\theta}$, a Gaussian process prior with mean $m_{m}=1/2$ and covariance $C_{m}=10I$ is utilized. For the SGDP and the SDP, the strength of neighbouring relationships, $\tau$, has a prior distribution Beta$(1/2, 1/2)$, and $\lambda(\cdot)$ in \eqref{eq:similarity} is set to the identity function. In SDP, $\alpha\sim {\rm Gamma}(1,1)$ and $\beta=1/\alpha$. 
Because the performances of the GDP and SGDP are dependent on the prior distributions of $\alpha$ and $\beta$, we implemented the following two cases of prior distributions based on the guidelines given at the end of Section~\ref{sec:priors}:
\begin{align*}
&{\rm Prior \ 1}: \ \alpha\sim {\rm Gamma}(2,1), \ \ \ \beta\sim {\rm Beta}(5,1),\\
&{\rm Prior \ 2}: \ \alpha\sim {\rm Gamma}(5,1), \ \ \ \beta\sim {\rm Beta}(20,1).
\end{align*}
The mean and variance of $\alpha\beta$ in Prior~1 are $1.667$ and $1.508$, and those in Prior~2 are $4.762$ and $4.597$, respectively. As measures of performance of point estimates, obtained by minimizing the posterior expected variation of information \citep{wade2018bayesian}, we employ two widely used metrics to assess the clustering performance: the adjusted Rand index \citep{hubert1985comparing}, abbreviated as ARI, and the purity function \citep{manning2009introduction}, denoted by PF. These measures gauge the concordance between actual and predicted cluster allocations. Note that high values of both ARI and PF indicate a high clustering accuracy. Additionally, the accuracy of the mean function estimation of the regions is quantified using the root mean squared error (RMSE) after normalizing by the $\ell_2$ norm of the true mean function. All Bayesian methods employ burn-in and sampling periods of $16000$ and $4000$, respectively. Note that the effective sample sizes of the parameters of interest are at least $500$. For the SGDP and SDP models, which incorporate the permutation sampling step, the Metropolis-Hastings algorithm for updating the permutation yielded an average acceptance rate of approximately 68.6\% and 64.3\% across the simulation studies.

\subsection{Result}

\begin{table}[t]
\centering
\caption{Adjusted Rand index (ARI), purity function (PF), and the root mean squared error (RMSE) for SDP, SGDP, and GDP with the different prior distributions.\label{tab:sim1}}
\begin{tabular}{llcccccc}
\toprule
\multicolumn{7}{c}{Case~1}\\
\midrule
SNR & Metric & \multicolumn{2}{c}{SGDP} & \multicolumn{2}{c}{GDP} & \multicolumn{1}{c}{SDP} &  \\ 
\cmidrule(lr){3-4} \cmidrule(lr){5-6} \cmidrule(lr){7-7}
       &        & Prior 1 & Prior 2 & Prior 1 & Prior 2 & - &  \\ \midrule
       & ARI     & 0.852 & 0.848 & 0.834 & 0.845 & 0.845 \\
High   & PF      & 0.858 & 0.861 & 0.856 & 0.841 & 0.846 \\ 
       & RMSE    & 0.103 & 0.098 & 0.106 & 0.113 & 0.114  \\
\midrule
       & ARI     & 0.647 & 0.675 & 0.636 & 0.649 & 0.637 \\
Low    & PF      & 0.665 & 0.698 & 0.644 & 0.660 & 0.650 \\ 
       & RMSE    & 0.103 & 0.100 & 0.115 & 0.117 & 0.107  \\
\toprule
\multicolumn{7}{c}{Case~2}\\
\midrule
SNR & Metric & \multicolumn{2}{c}{SGDP} & \multicolumn{2}{c}{GDP} & \multicolumn{1}{c}{SDP} &  \\ 
\cmidrule(lr){3-4} \cmidrule(lr){5-6} \cmidrule(lr){7-7}
     &        & Prior 1 & Prior 2 & Prior 1 & Prior 2 &  \\ \midrule
     & ARI    & 0.846   & 0.842   & 0.837   & 0.834   & 0.841 \\
High  & PF     & 0.844   & 0.863   & 0.865   & 0.829   & 0.837 \\ 
     & RMSE   & 0.112   & 0.101   & 0.108   & 0.143   & 0.140 \\ 
\midrule
     & ARI    & 0.650   & 0.646   & 0.611   & 0.625   & 0.638 \\
Low  & PF     & 0.661   & 0.692   & 0.635   & 0.647   & 0.643 \\ 
     & RMSE   & 0.102   & 0.103   & 0.138   & 0.124   & 0.110 \\
\bottomrule
\end{tabular}
\end{table}

The results presented in Table~\ref{tab:sim1}, which presents the average of the evaluation metrics across the $50$ datasets, provide an overview of the clustering performance. The findings demonstrate that the SGDP consistently outperforms the other methods, regardless of the data-generating scenario. Specifically, the SGDP achieves the highest ARI values, indicating its superior clustering accuracy and consistency. This result suggests that the SGDP can effectively identify and cluster similar data points, even in the presence of noise and variability. Similarly, the SGDP attains the highest scores in terms of PF, particularly in low SNR scenarios. While SDP tends to discern minor differences and finely categorize them, the SGDP mitigates this tendency, striking a balance between capturing meaningful distinctions and avoiding over-segmentation. These results suggest that the SGDP is proficient in clustering data across different clusters and accurately capturing the characteristics of each cluster. Furthermore, the SGDP exhibits the lowest RMSE values, which highlight its effectiveness in precise mean function estimation. Thus, the proposed SGDP is a robust and reliable choice for a wide range of clustering applications.

\begin{table}[t]
\centering
\caption{Posterior summaries of $(\alpha, \beta)$. \label{tab:sim2}}

\begin{tabular}{llcccc}
\toprule
\multicolumn{6}{c}{Case~1}\\
\midrule
SNR & Percentile & \multicolumn{2}{c}{SGDP} & \multicolumn{2}{c}{GDP} \\ 
\cmidrule(lr){3-4} \cmidrule(lr){5-6} 
     &           & Prior 1 & Prior 2 & Prior 1 & Prior 2 \\
\midrule
     & 97.5$\%$   & (5.52, 0.22) & (4.29, 0.63) & (4.65, 0.48) & (4.75, 0.64)  \\
High & 50$\%$     & (5.46, 0.20) & (4.25, 0.55) & (4.55, 0.31) & (4.68, 0.60)  \\ 
     & 2.5$\%$    & (5.36, 0.19) & (4.22, 0.50) & (4.50, 0.25) & (4.54, 0.53)  \\
\midrule
     & 97.5$\%$   & (5.58, 0.24) & (5.14, 0.56) & (4.86, 0.66) & (4.95, 0.79)  \\
Low  & 50$\%$     & (5.53, 0.21) & (5.09, 0.53) & (4.83, 0.59) & (4.90, 0.71)  \\ 
     & 2.5$\%$    & (5.47, 0.19) & (5.04, 0.42) & (4.76, 0.47) & (4.86, 0.62)  \\
\toprule
\multicolumn{6}{c}{Case~2}\\
\midrule
SNR & Percentile & \multicolumn{2}{c}{SGDP} & \multicolumn{2}{c}{GDP} \\ 
\cmidrule(lr){3-4} \cmidrule(lr){5-6} 
     &           & Prior 1 & Prior 2 & Prior 1 & Prior 2 \\
\midrule
     & 97.5$\%$   & (5.48, 0.23) & (4.34, 0.61) & (4.61, 0.42) & (4.81, 0.67)  \\
High & 50$\%$     & (5.43, 0.22) & (4.28, 0.56) & (4.56, 0.40) & (4.66, 0.65)  \\ 
     & 2.5$\%$    & (5.26, 0.29) & (4.24, 0.53) & (4.52, 0.37) & (4.61, 0.62)  \\
\midrule
     & 97.5$\%$   & (5.66, 0.27) & (5.15, 0.60) & (4.99, 0.62) & (5.01, 0.79)  \\
Low  & 50$\%$     & (5.61, 0.24) & (5.11, 0.59) & (4.80, 0.62) & (4.96, 0.75)  \\ 
     & 2.5$\%$    & (5.56, 0.23) & (5.02, 0.56) & (4.81, 0.60) & (4.91, 0.72)  \\
\bottomrule
\end{tabular}%

\end{table}

Table~\ref{tab:sim2} summarizes the posterior distributions for the parameters $\alpha$ and $\beta$. The choice of prior distribution influences the posterior distribution of $\alpha\beta$ and attempts to increase $\alpha\beta$ in the prior distribution results in large posterior values. The posterior distributions of $\alpha$ and $\beta$ were significantly larger than 1, and this tendency is more pronounced when Prior~2 is employed. This result, in conjunction with the findings presented in Table~\ref{tab:sim1}, suggests that higher $\alpha\beta$ values are instrumental in preventing the formation of excessive clusters, particularly in low SNR scenarios. This is an important consideration, as over-clustering can lead to a loss of interpretability of the results.

\section{Application: Clustering hourly population data in Tokyo}\label{sec:app}
In this section, we examine a case study involving population data from Tokyo, Japan, to investigate the efficacy of the proposed methodology. Initially, the characteristics of the population data are described. Subsequently, we explain models that capture the distinct features of the data. Lastly, we discuss the clustering results, focusing on the SGDP parameters and spatial correlation.

\subsection{Hourly population data}
The dataset under examination is population data collected by NTT Docomo Inc., the predominant mobile company in Japan with about $82$ million users across the country. The company leverages user data to estimate the number of mobile phone users among all mobile carriers in each region. Based on observations and mobile phone penetration rates, the population of each region is estimated with a high degree of accuracy \citep{oyabu2013evaluating}. In this study, we considered the seven special wards of Tokyo's metropolitan area, with each mesh defined as a $500 \text{m}$ mesh, resulting in $n=452$ such units (refer to Figure~\ref{fig:districts}). Hourly population data was collected within each mesh over $T=30$ days, commencing on January $21$, 2019. Recognizing that the variation in population flows reflected the unique characteristics of each region, we standardised the scales among the regions as follows:
\begin{equation*}
    \frac{y_{it}(x)}{\sqrt{\sum_{i,t}y_{it}^2(x)/n/T}}.
\end{equation*}

\subsection{Model}
Let $y_{it}(x)$ be the observed functional data (population) in the area (mesh) $i\in\{1,2,\ldots,n\}$ and at time $t\in \{1,2,\ldots,T\}$. In this application, $t$ and $x\in\mX$ represent a day and a time within the day, respectively. Analysing functions observed across various points in time and space necessitates a methodology adept at capturing regions' or periods' specific attributes by identifying patterns.

Although our initial focus was on clustering based only on districts, we recognize that temporal data structures often contain information that should not be ignored and should be incorporated to glean deeper insights. This issue is addressed in the current section. We consider extensions that allow period-to-period cluster changes as follows:
\begin{equation}\label{mixture-st}
\begin{split}\centering
&y_{it}(x)\mid\mu_{it}(x)\sim \GP(\mu_{it}(x), C_y),\\
& \mu_{it}(x)= \sum_{\ell=1}^{M} w_{t\ell}\left\{\sum_{j=1}^{K_\ell} \theta_{j\ell}(x)I(z_{i\ell}=j)\right\},\\
& \theta_{j\ell}(x) \sim \GP(m_\theta^{(\ell)},C_\theta^{(\ell)}),\ \ \ j=1,\ldots,K_{\ell}, \ \ \ \ell=1,\ldots,M, 
\end{split}
\end{equation}
where $M$ is the number of periods, $z_{i1},\ldots,z_{iM}$ represent the cluster membership variables, $w_{t\ell}$ denotes the period indicator, and $(m_\theta^{(\ell)},C_\theta^{(\ell)})$ indicates the Gaussian process parameters for each period $\ell$ in each cluster $j$. If $T=M=1$, \eqref{mixture-st} coincides with \eqref{mixture}. Because the clusters depend on periods, the membership variables are given for each period; that is, we considered the prior distribution $(z_{1\ell},\ldots,z_{n\ell})\sim {\rm SGDP}(\alpha_{\ell},\beta_{\ell},\tau_{\ell})$ for $\ell=1,\ldots,M$. Furthermore, we employed RBF kernels to model the covariance matrices $C_y$ and $C_{\theta}^{(\ell)}$ for $\ell=1,\ldots,M$.

\begin{figure}[tb]
    \centering
    \includegraphics[width=0.5\linewidth]{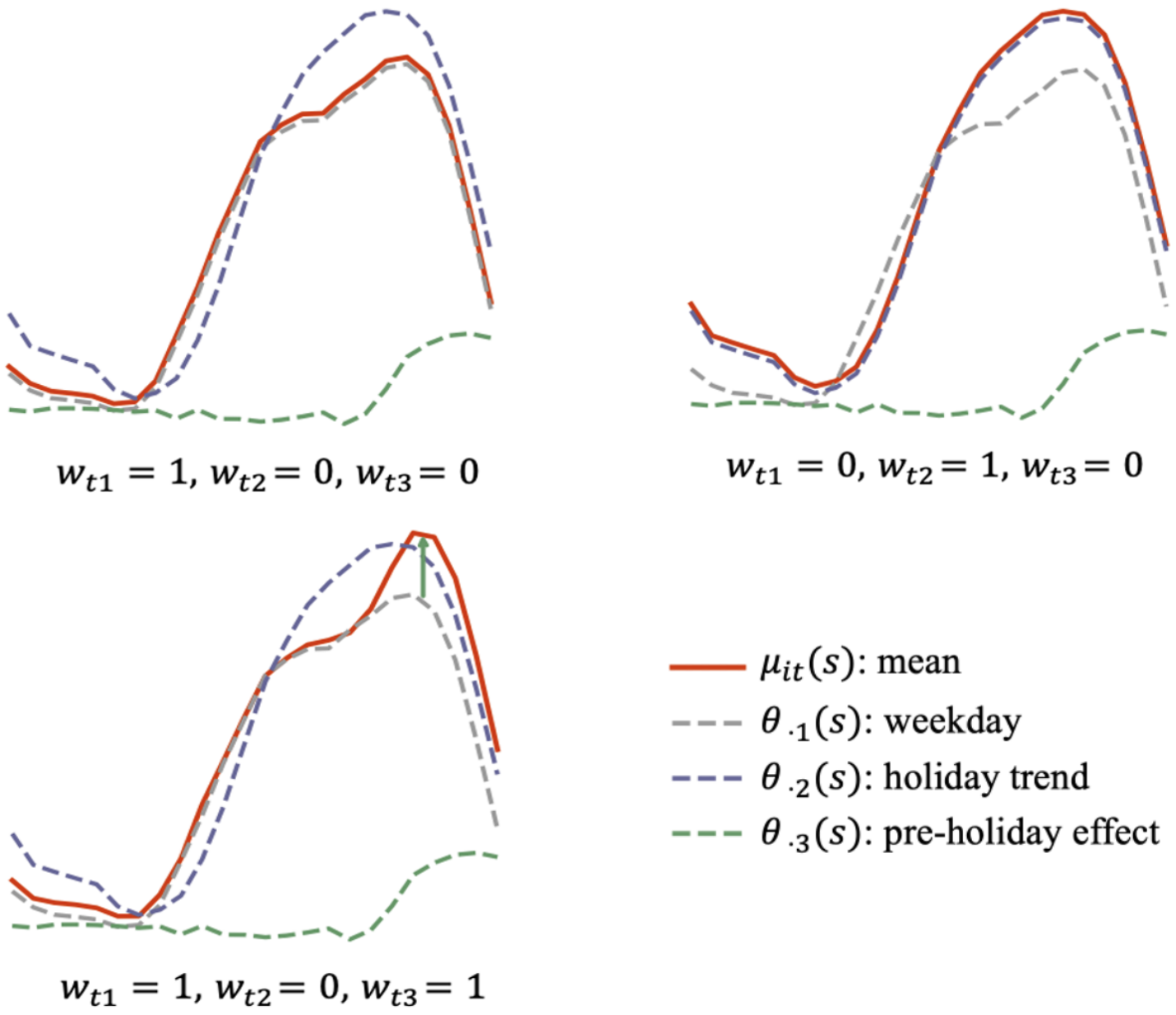}
    \caption{Weekday trend (top--left), holiday trend (top--right), and trend for the day before a holiday with the upward arrow indicating the pre-holiday effect (bottom--left).}
    \label{fig:temporal}
\end{figure}

The notable distinction from \eqref{mixture} lies in the introduction of $w_{t\ell}$, which we elucidate here. Consider a scenario where $T=2$, with day $1$ being a weekday $(w_{t1}=1,\ w_{t2}=0)$, and day $2$ a holiday $(w_{t1}=0,\ w_{t2}=1)$. The population flows on these days are markedly different between weekdays and holidays; hence, they are likely to exhibit varied clustering patterns. Additionally, Fridays may experience unique nighttime population increases in downtown areas compared to other weekdays, in which case $w_{t1}=1,\ w_{t2}=0,\ w_{t3}=1$.
The grey line in Figure~\ref{fig:temporal} represents the weekday trend, the blue line represents the holiday trend, and the green line represents the pre-holiday effect. For instance, on Fridays, the observed data combines the weekday trend and the green line's effect (plus noise).
This detailed temporal structure aids in deepening our understanding of spatial patterns. In our analysis, we set the number of periods $M$ to three, with the weekday, holiday, and pre-holiday indicators for each period being $w_{t1}$, $w_{t2}$, and $w_{t3}$, respectively.

\subsection{Implementation}
We performed Bayesian inference based on the model reported in \eqref{mixture-st}. The prior distributions for the range parameters of the covariance kernels, $\phi_y$ and $\phi_{\theta,\ell}$, along with the scale parameters, $\eta_y$ and $\eta_{\theta,\ell}$, are all modeled using an IG$(1/2, 1/2)$ distribution for $\ell=1,\ldots,M$. For $m_{\theta}^{(\ell)}$, we employ a Gaussian process prior with mean $m_{m}^{(\ell)}=1/2$ and covariance $C_{m}^{(\ell)}=10I$ for each $\ell$. $\tau$ follows a Beta$(1/2, 1/2)$ prior distribution.
The prior for $\alpha_{\ell}$ is set as Gamma$(5,1)$ and the prior for $\beta_{\ell}$ as Beta$(10,1)$ to prevent generating excessive clusters for each $\ell$, implying that the prior mean and variance of $\alpha_\ell\beta_\ell$ are equal to $4.545$ and $4.338$, respectively. Alongside the model in \eqref{mixture-st}, we also implemented the SDP, whereby $\alpha_{\ell}\sim {\rm Gamma}(1,1)$ and $\beta_{\ell}=1/\alpha_{\ell}$ for all $\ell$, and GDP, where $\tau_{\ell}=1$ in the model \eqref{mixture-st}. All the methods are implemented using the MCMC technique with a burn-in period of $16000$ and a sampling period of $4000$. The representative value is calculated by minimizing the variation of information~\citep{wade2018bayesian}. In this application, the Metropolis-Hastings step for sampling permutations within the SGDP and SDP models (and the model in \eqref{mixture-st} when applicable) had an average acceptance rate of approximately 58.0\% and 61.9\%.

\begin{table}[t]
\centering
\caption{Posterior summaries of $(\alpha_{\ell}, \beta_{\ell})$ in the SGDP. \label{tab:sim3} }
\begin{tabular}{lccc}
\toprule
 Percentile & \multicolumn{3}{c}{Period}   \\ 
\cmidrule(lr){2-4} 
        &  1 & 2 & 3 \\
\midrule
 97.5$\%$   & (2.10, 0.96) & (1.06, 0.97) & (1.40, 0.96)   \\
 50$\%$     & (1.97, 0.94) & (1.05, 0.96) & (1.36, 0.95)  \\ 
 2.5$\%$    & (1.90, 0.93) & (1.05, 0.94) & (1.31, 0.93)   \\
\bottomrule
\end{tabular}
\end{table}

\subsection{Effect of generalized parametrization}
\begin{figure}[t]
    \centering\includegraphics[width=0.6\linewidth]{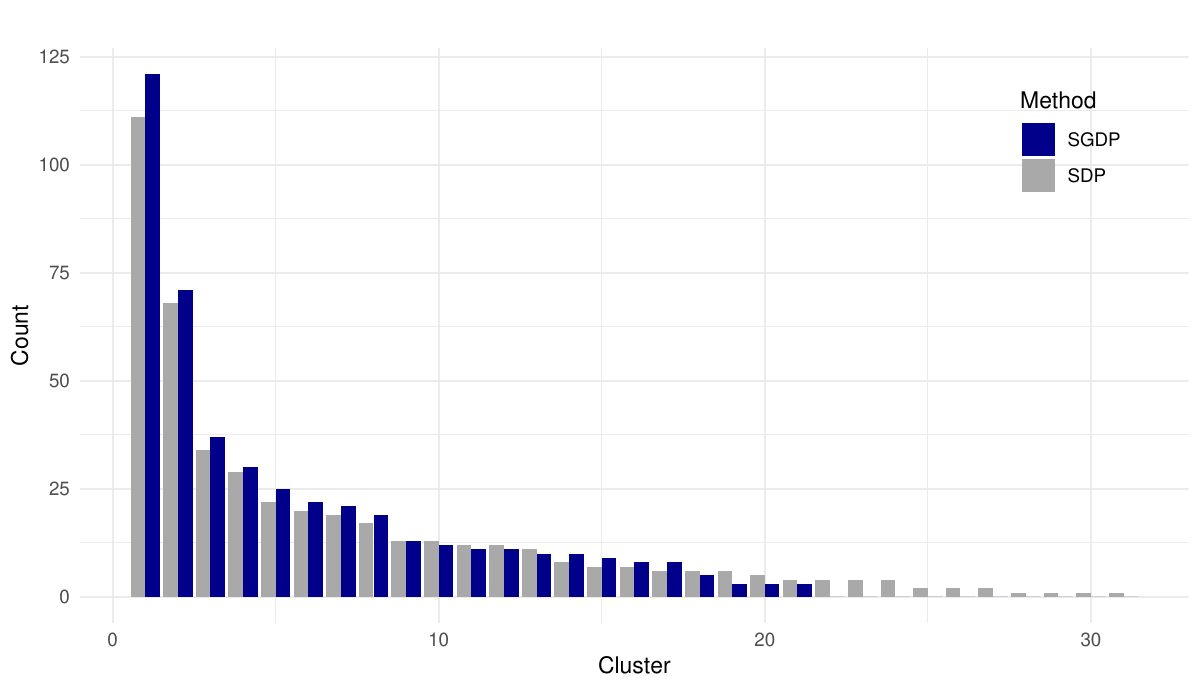}
    \caption{Distribution of the cluster sizes for weekday obtained from the SGDP and SDP. }
    \label{fig:dist30}
\end{figure}

\begin{figure}[t]
    \centering\includegraphics[width=0.6\linewidth]{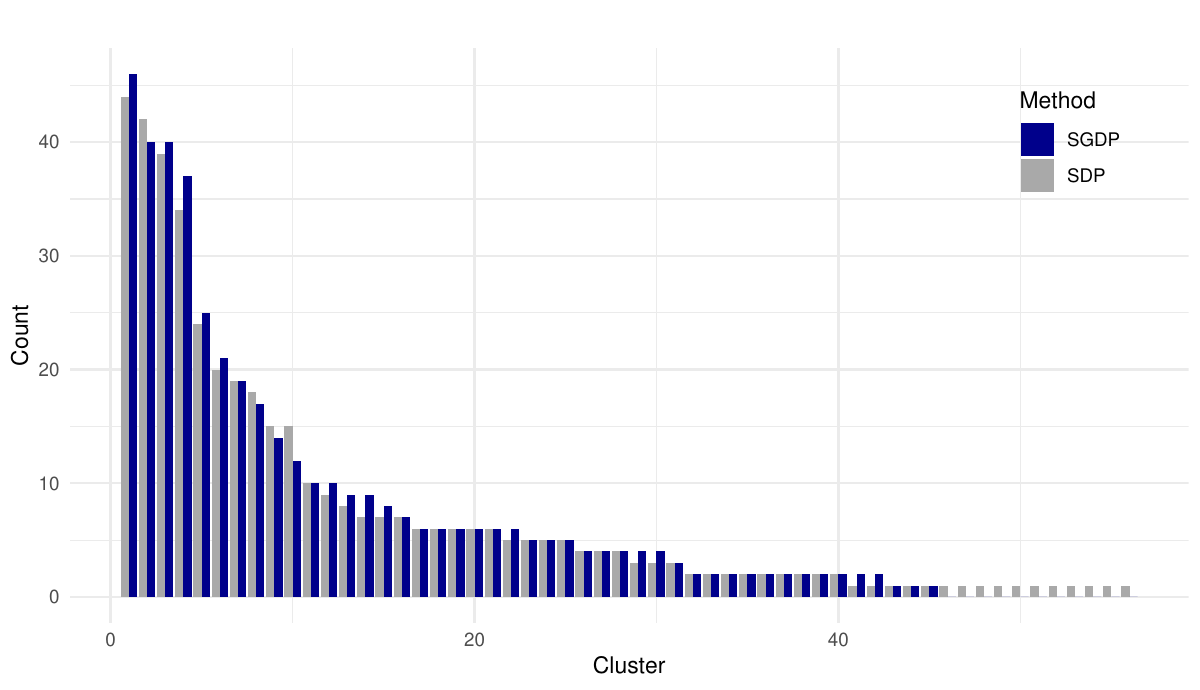}
    \caption{Distribution of the cluster sizes for holidays obtained from the SGDP and SDP. }
    \label{fig:dist2}
\end{figure}

\begin{figure}[t]
    \centering\includegraphics[width=0.6\linewidth]{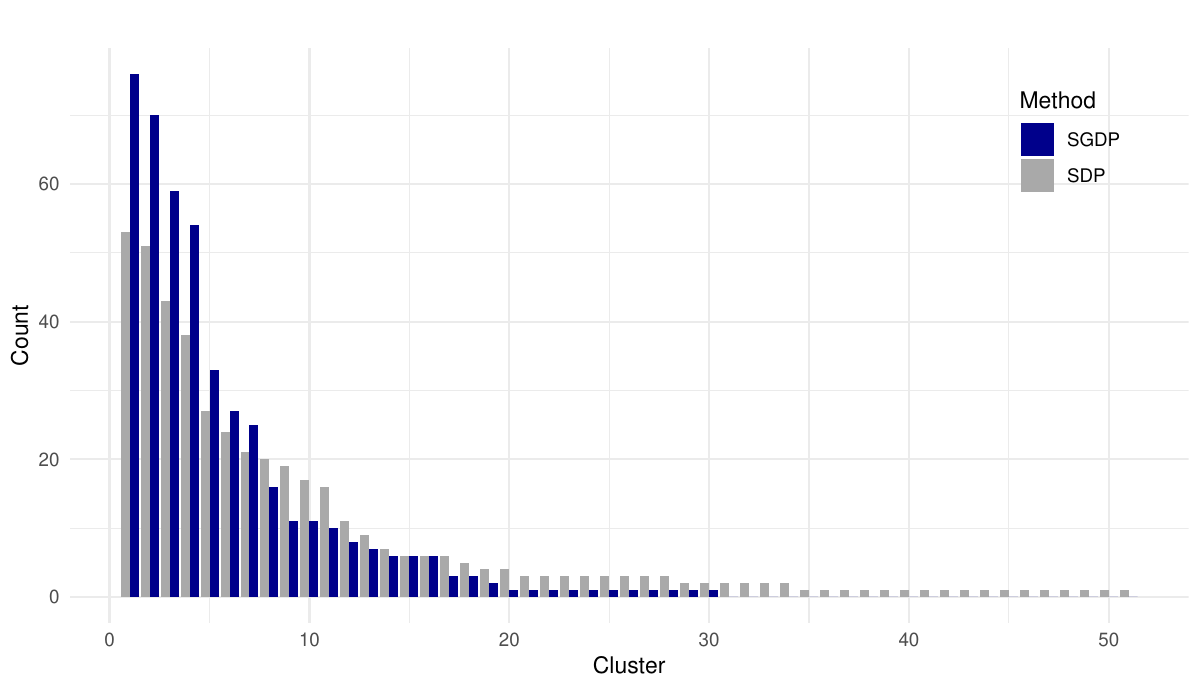}
    \caption{Distribution of the cluster sizes for the pre-holiday effect obtained from the SGDP and SDP. }
    \label{fig:dist3}
\end{figure}

First, we analyse the differences between the SGDP and SDP by examining the distribution of the cluster numbers. Introducing priors leads to posterior estimates congruent with the data and prior beliefs, thereby reducing cluster numbers. Table~\ref{tab:sim3} presents the posterior summaries of $(\alpha_{\ell}, \beta_{\ell})$ in the SGDP. The posterior values of $\alpha_{\ell}\beta_{\ell}$ exceed $1$ for each $\ell$, and the results tend to avoid redundant clusters. The distributions of cluster size for these two methods in weekdays, holidays, and pre-holidays are illustrated in Figures~\ref{fig:dist30},~\ref{fig:dist2}~and~\ref{fig:dist3}, respectively. From the weekday analysis in Figures~\ref{fig:dist30}, the SDP exhibits $8$ out of $29$ clusters consisting of only a single item, indicating a long tail to the right. Conversely, the SGDP shows a heavier left-side mass, resulting in a more attenuated tail. A similar pattern can be observed consistently across other periods. These findings suggest that owing to its flexible parametrization, the SGDP can effectively address the issue of over-clustering associated with increased dimensionality.


\begin{figure}[t]
\centering
\begin{tabular}{cc}
    \includegraphics[page=1,width=0.45\linewidth]{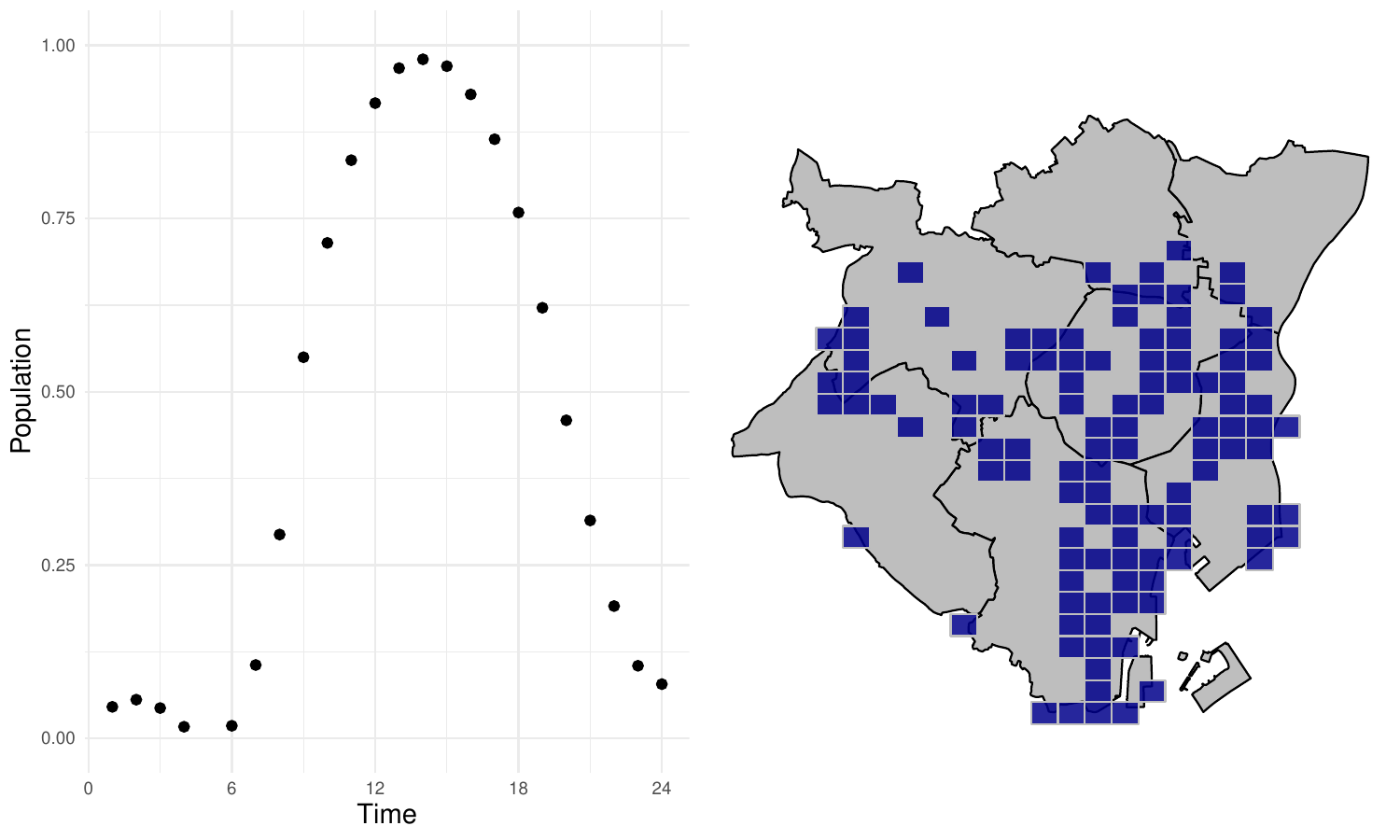} &
    \includegraphics[page=1,width=0.45\linewidth]{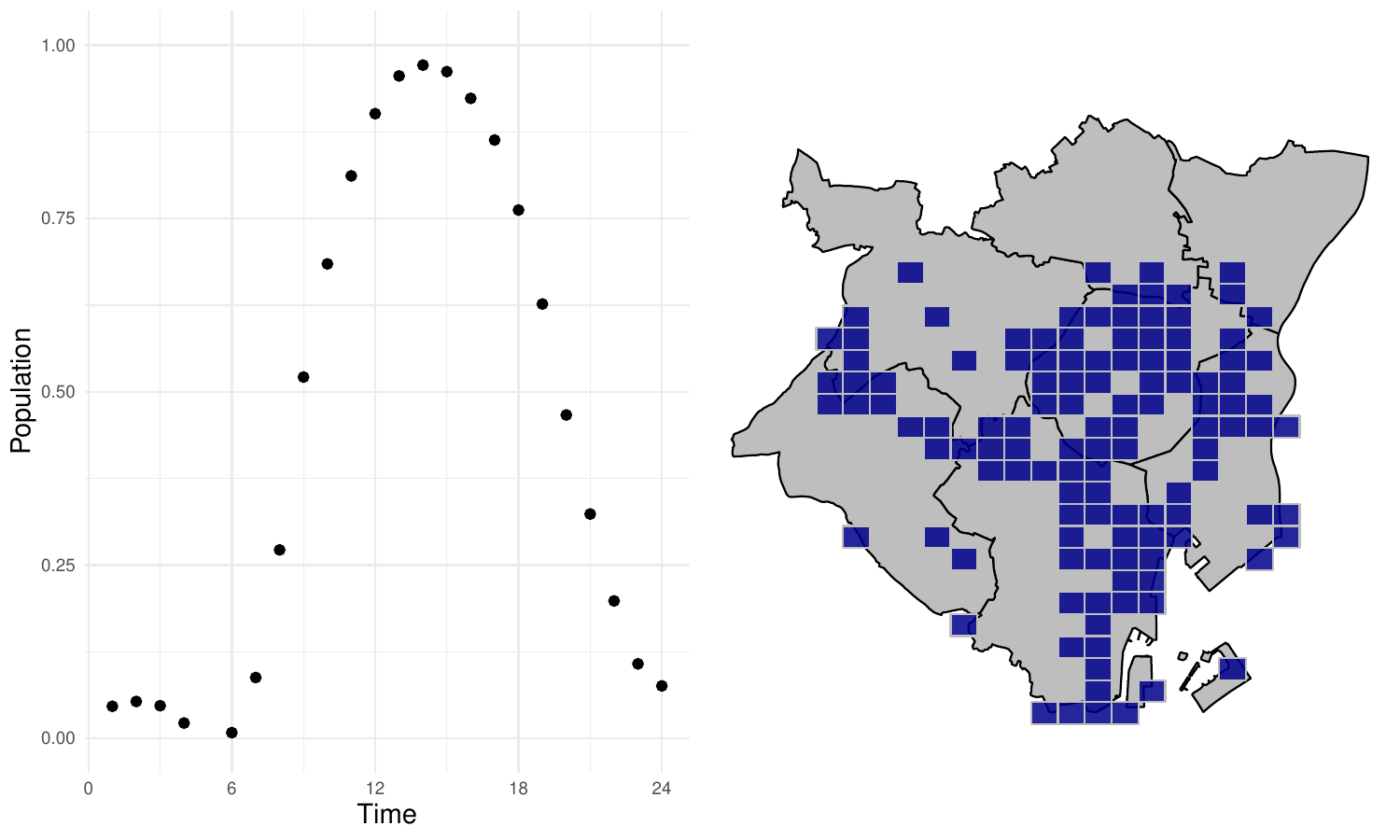} \\
    \includegraphics[page=1,width=0.45\linewidth]{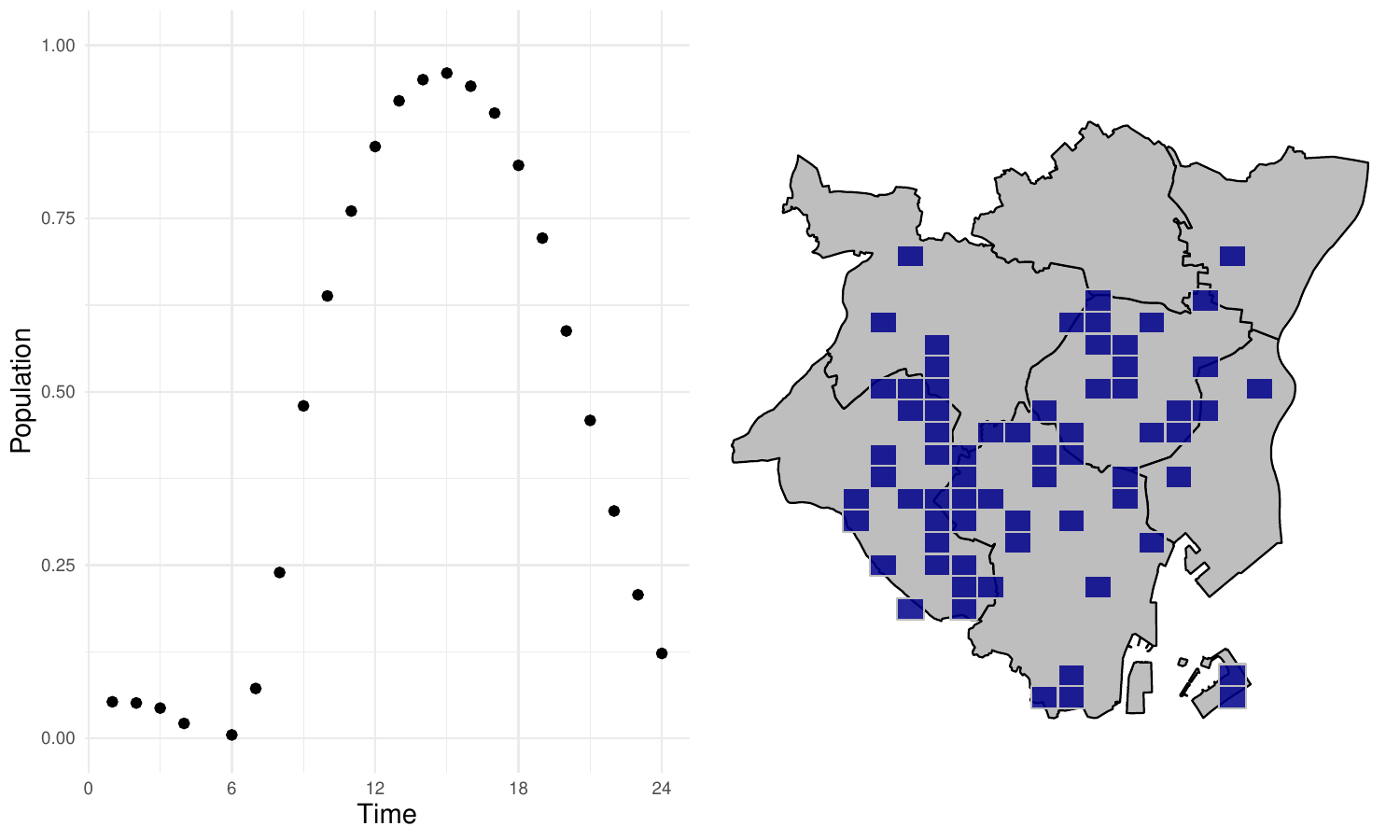} &
    \includegraphics[page=1,width=0.45\linewidth]{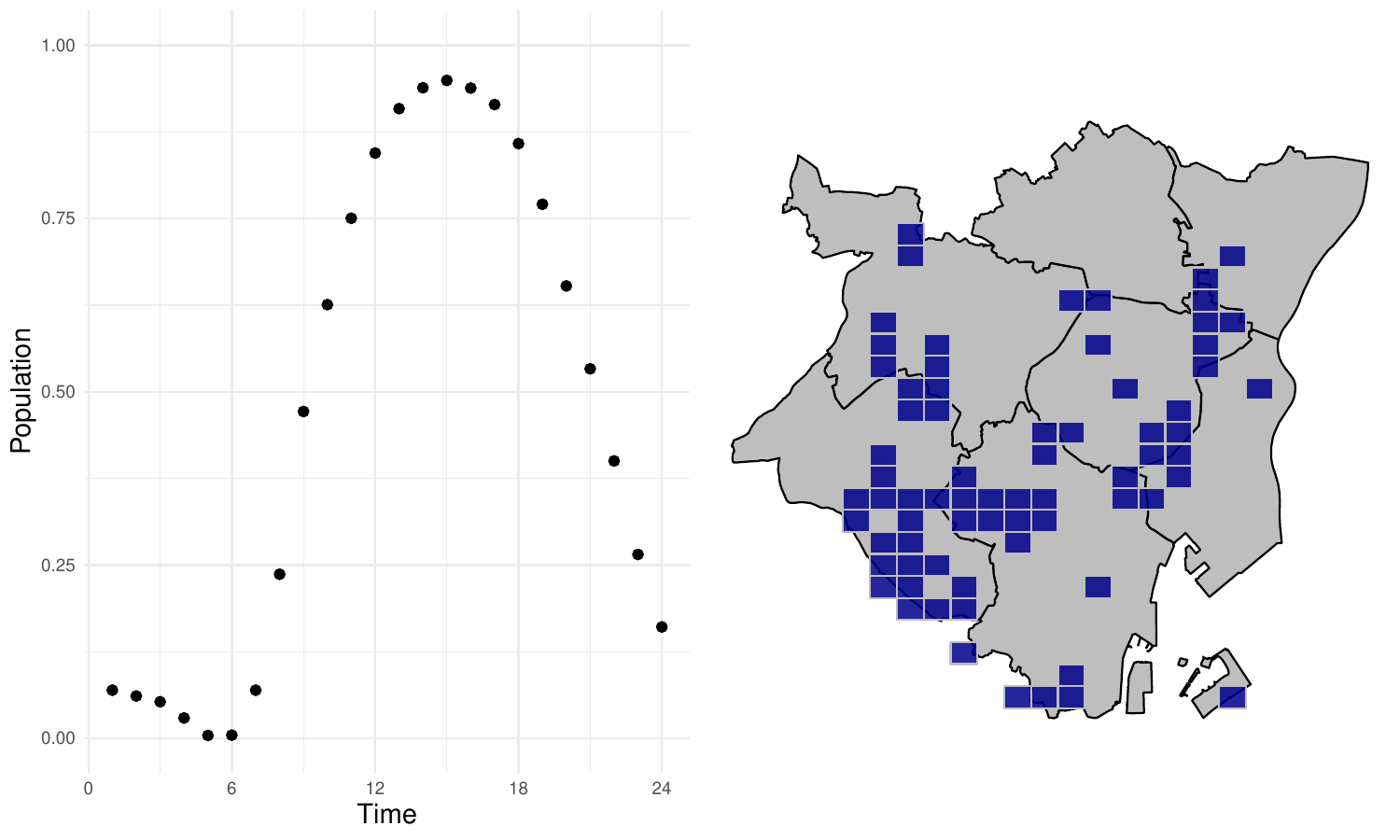} \\
    \includegraphics[page=1,width=0.45\linewidth]{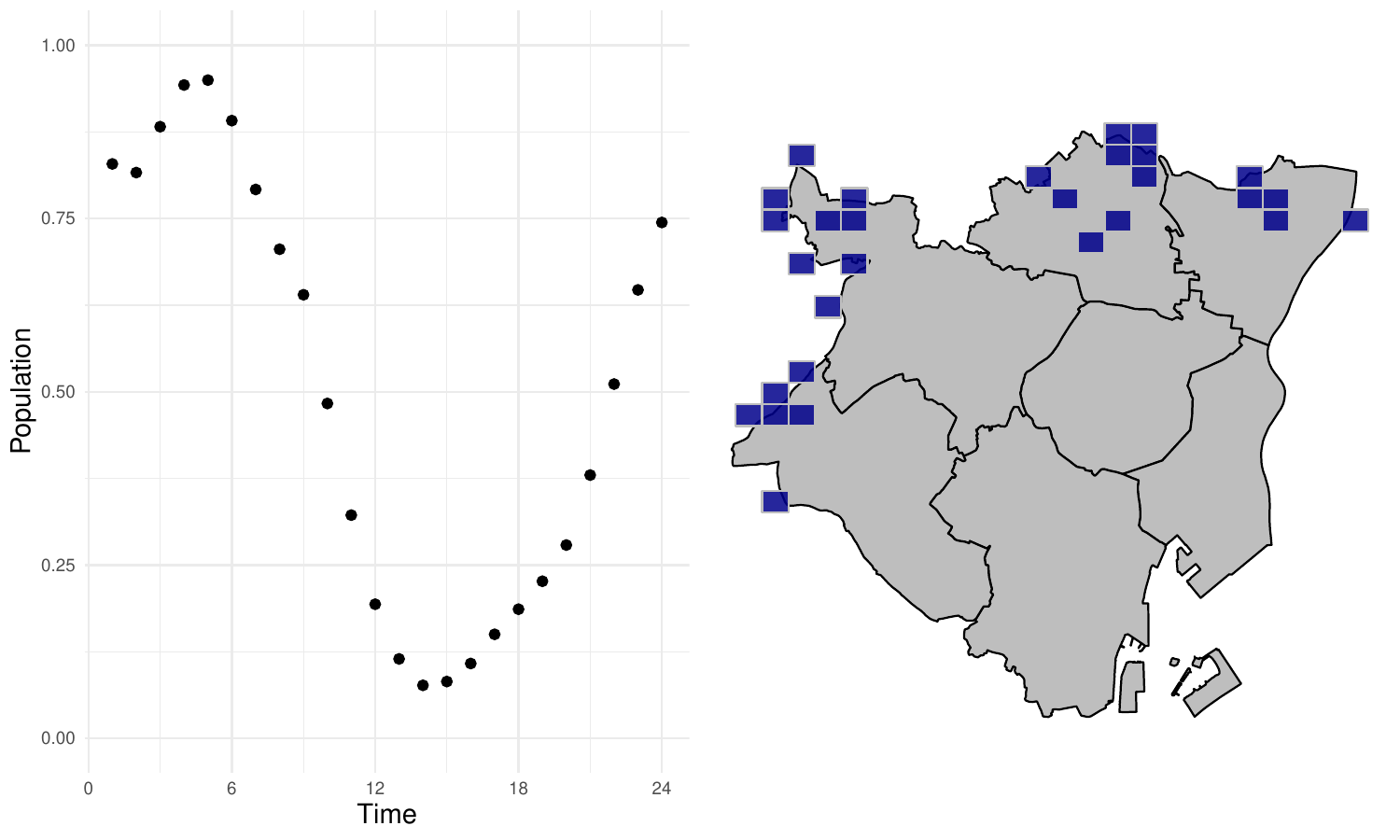} &
    \includegraphics[page=1,width=0.45\linewidth]{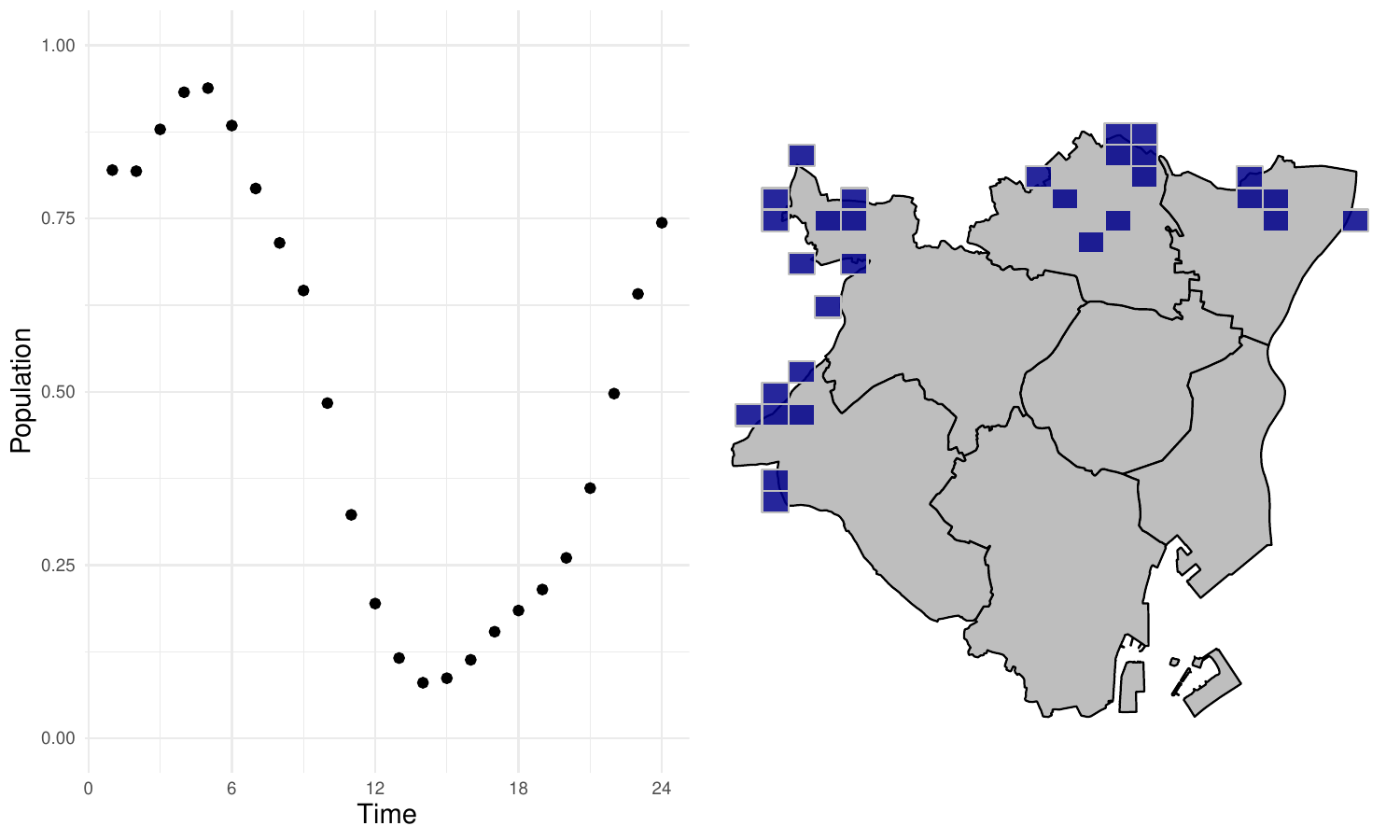} \\
\end{tabular}\caption{Three largest clusters on weekdays using the SDP (left) and SGDP (right). Each row represents a different type of area: the top row is the office area, the middle row is the downtown, and the bottom row is the residential area.}
\label{tab:cluster_plot}
\end{figure}

Next, we examined the clusters on weekdays ($\ell=1$) formed by the SGDP and SDP methods. Figure~\ref{tab:cluster_plot} depicts the three largest clusters on weekdays. The SDP-generated clusters are in the left column, while those generated by the SGDP are in the right column. The first row displays the weekday population flows in the business area and their corresponding mapped areas. The second and third rows represent downtown and residential areas, respectively. Evidently, the daytime population increases in the business and downtown areas, whereas it decreases in the residential areas, implying movement from residential to other areas for work or shopping, aligning with findings from other urban case studies \citep[e.g.,][]{xie2021revealing}. This result indicates that the two methods reflecting spatial information adequately discover regional characteristics.

\begin{figure}[t]
\centering
\begin{tabular}{cc}
    \includegraphics[page=1,width=0.45\linewidth]{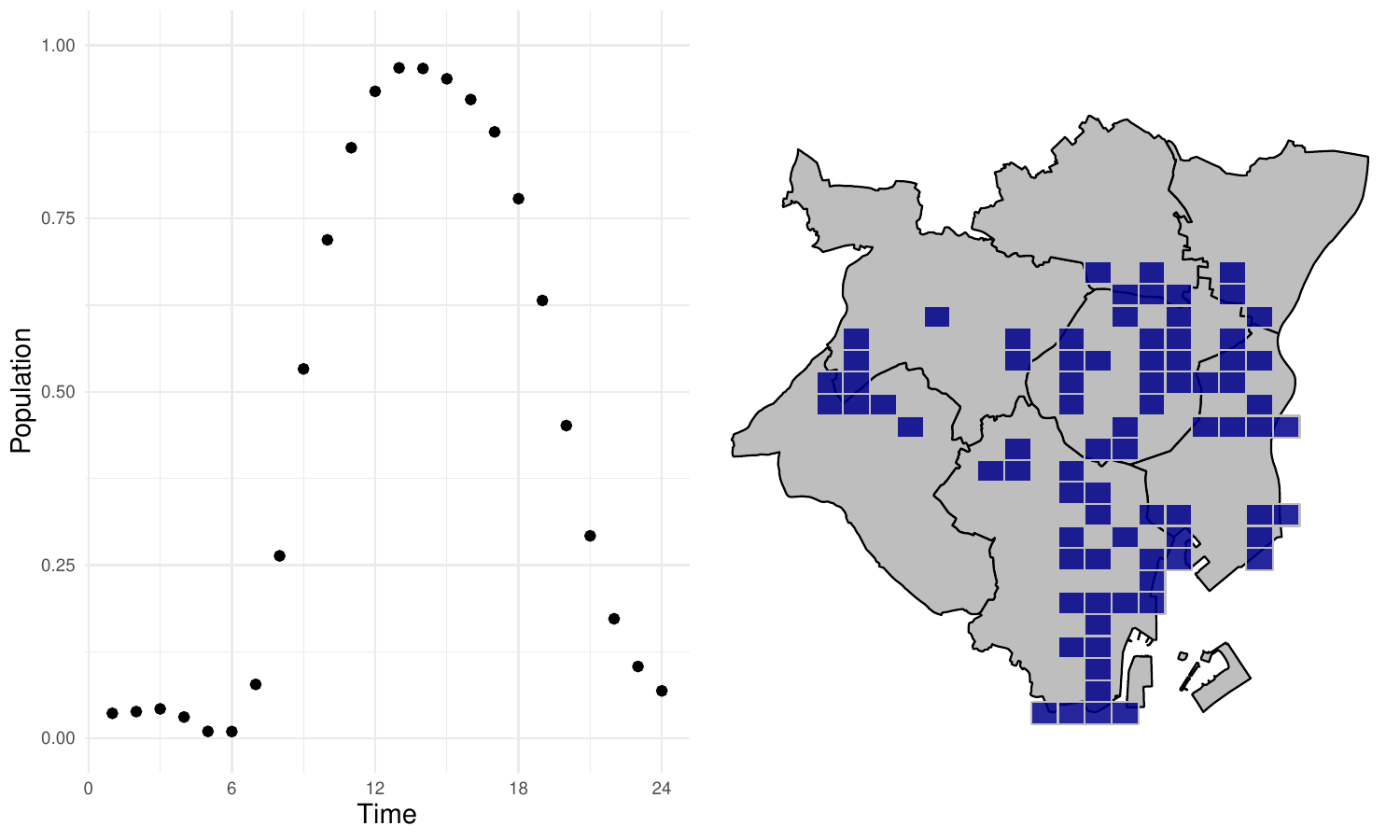} &
    \includegraphics[page=1,width=0.45\linewidth]{GDP_office.pdf} \\
    \includegraphics[page=1,width=0.45\linewidth]{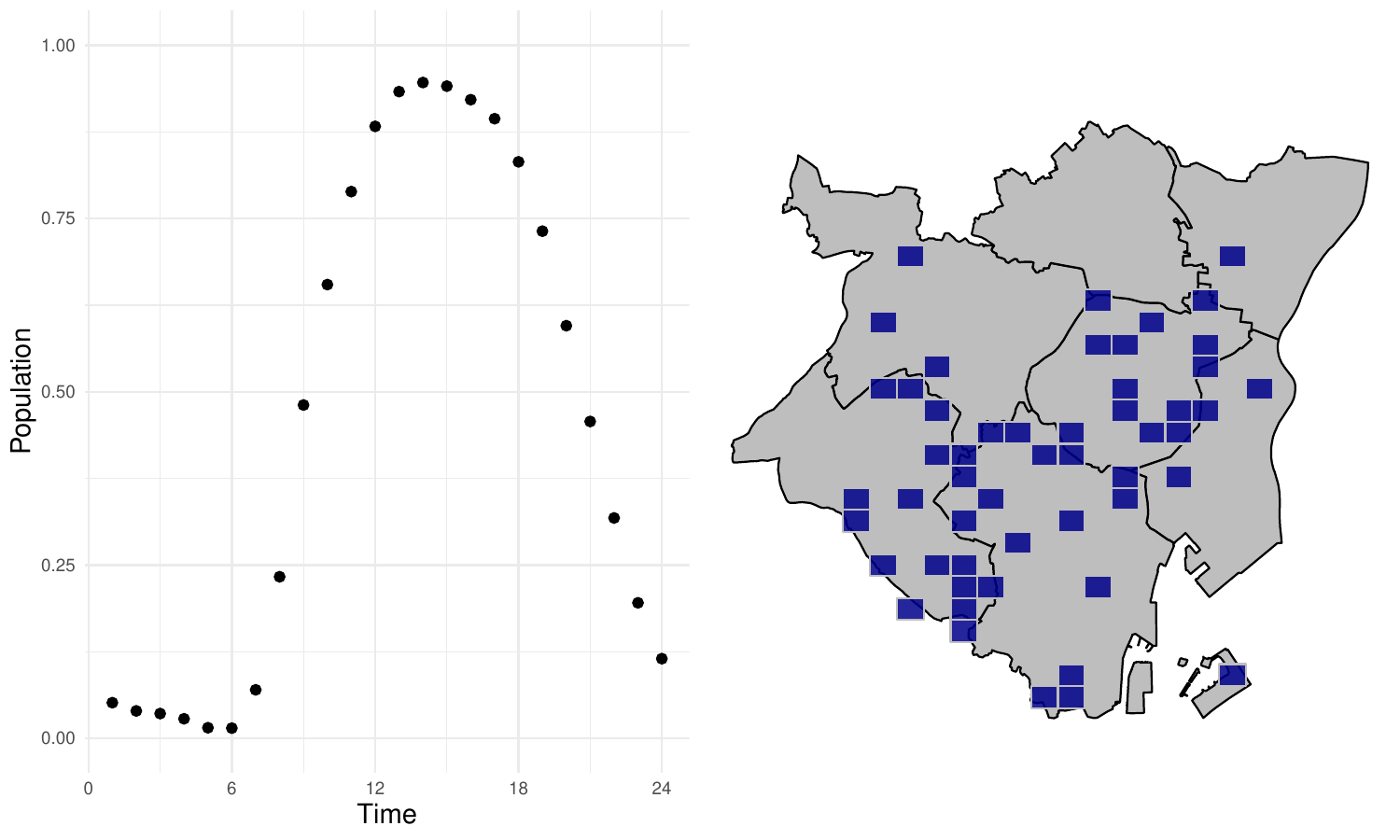} &
    \includegraphics[page=1,width=0.45\linewidth]{GDP_downtown.pdf} \\
    \includegraphics[page=1,width=0.45\linewidth]{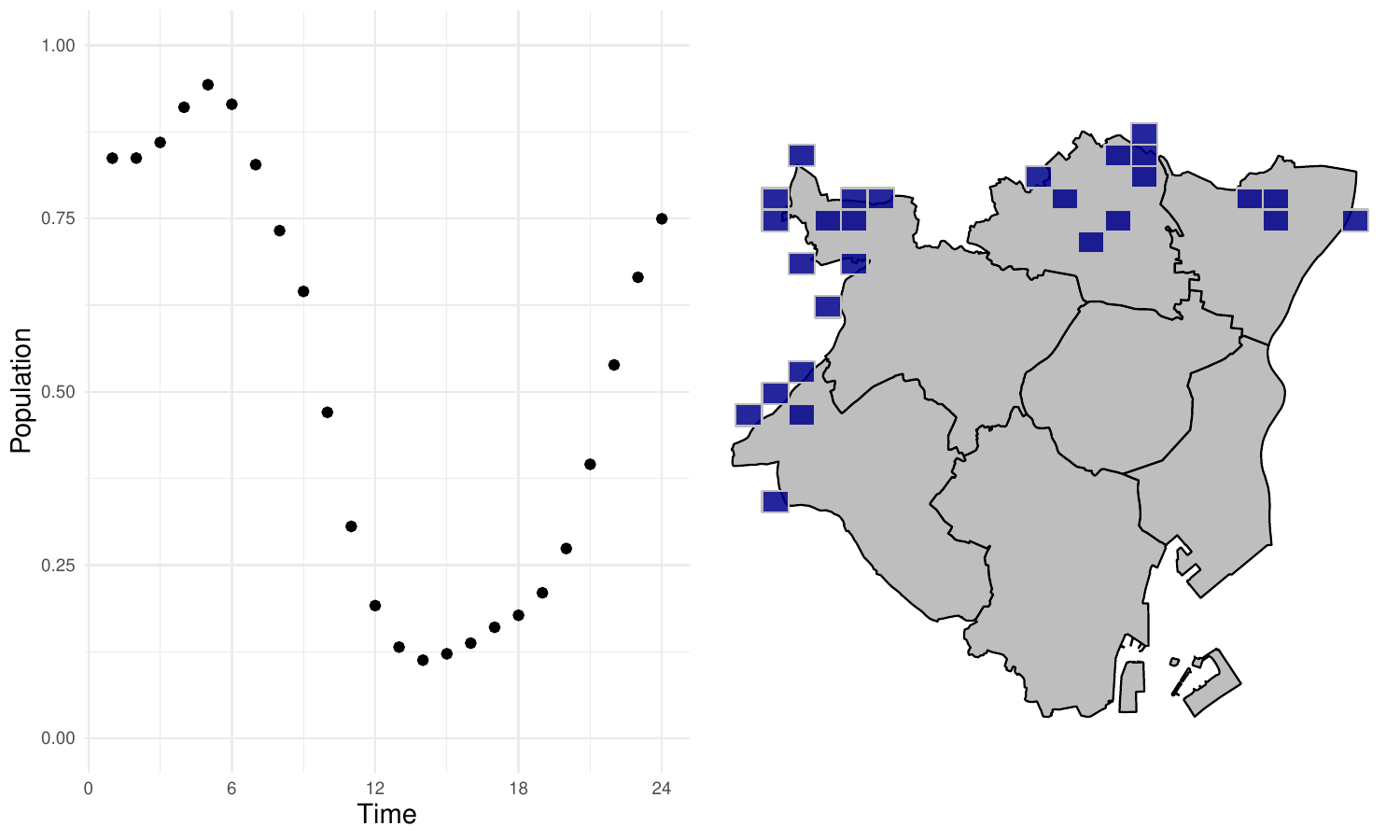} &
    \includegraphics[page=1,width=0.45\linewidth]{GDP_residential.pdf} \\
\end{tabular}
\caption{Three largest clusters on weekdays using simple GDP (left) and SGDP (right). Each row represents a different type of area: the top row indicates the office area, the middle row denotes the downtown area, and the bottom row highlights the residential area.}
\label{tab:ns_plot}
\end{figure}

\begin{figure}[t]
    \centering
    \includegraphics[width=0.6\linewidth]{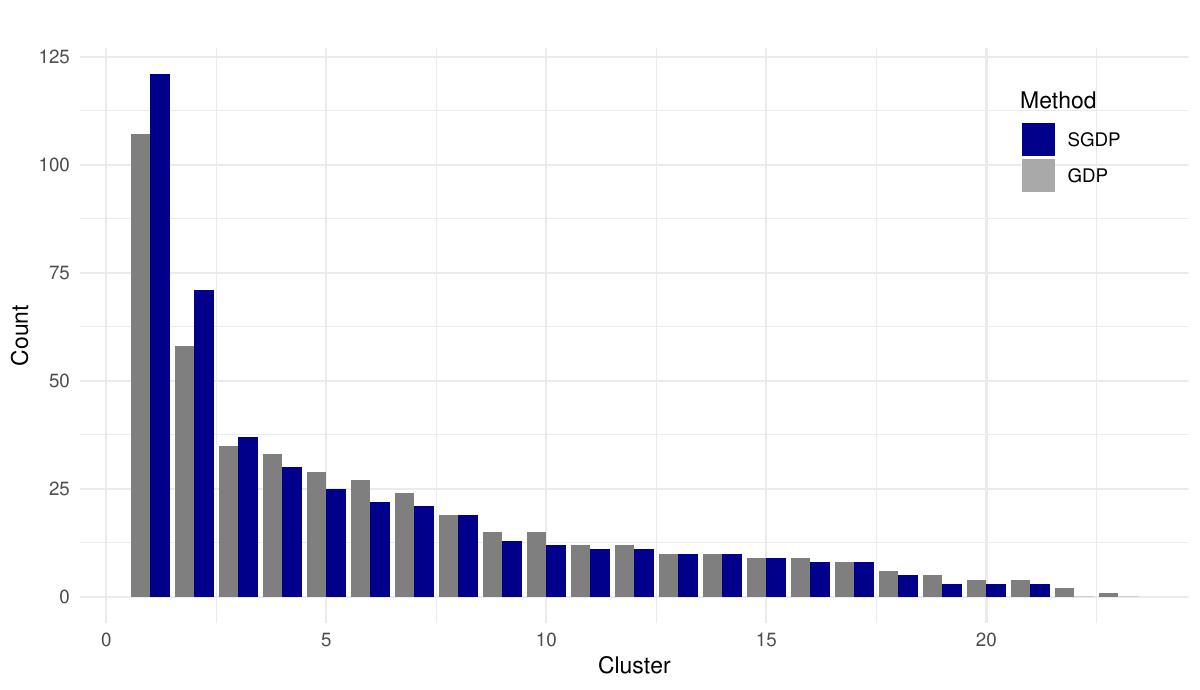}
    \caption{Distribution of the cluster sizes for weekdays obtained from the SGDP and GDP. }
    \label{fig:dist_ns}
\end{figure}
\begin{figure}[t]
    \centering\includegraphics[width=0.6\linewidth]{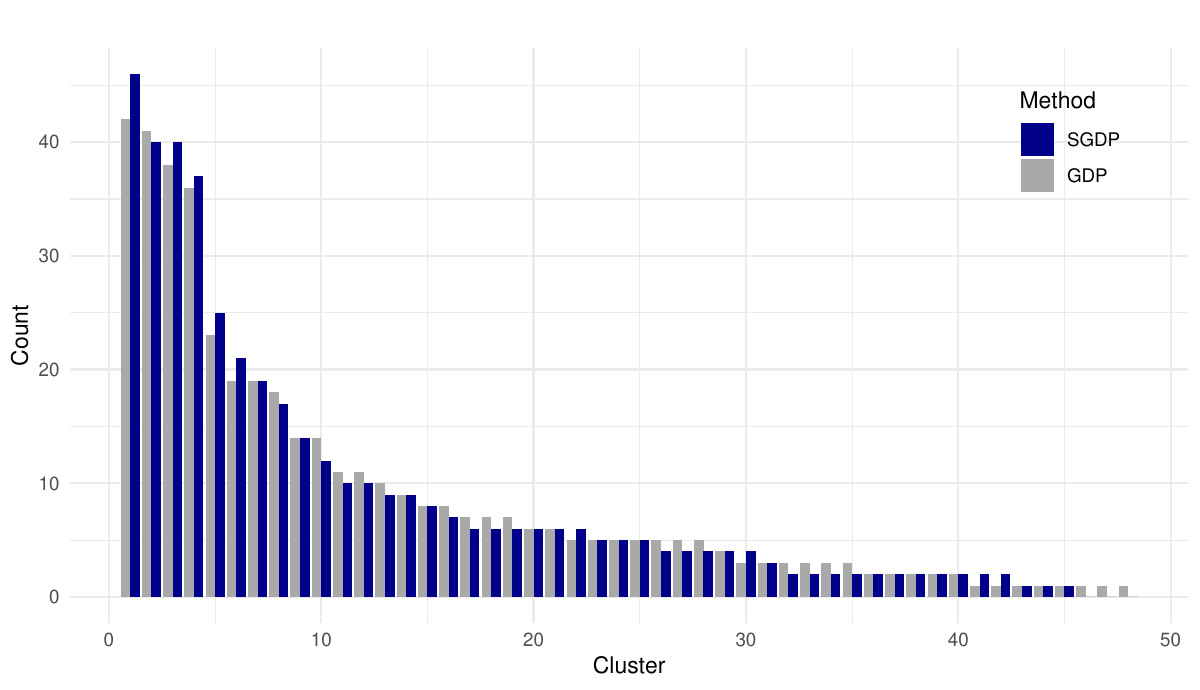}
    \caption{Distribution of the cluster sizes for holidays obtained from the SGDP and GDP. }
    \label{fig:dist_ns2}
\end{figure}
\begin{figure}[t]
    \centering\includegraphics[width=0.6\linewidth]{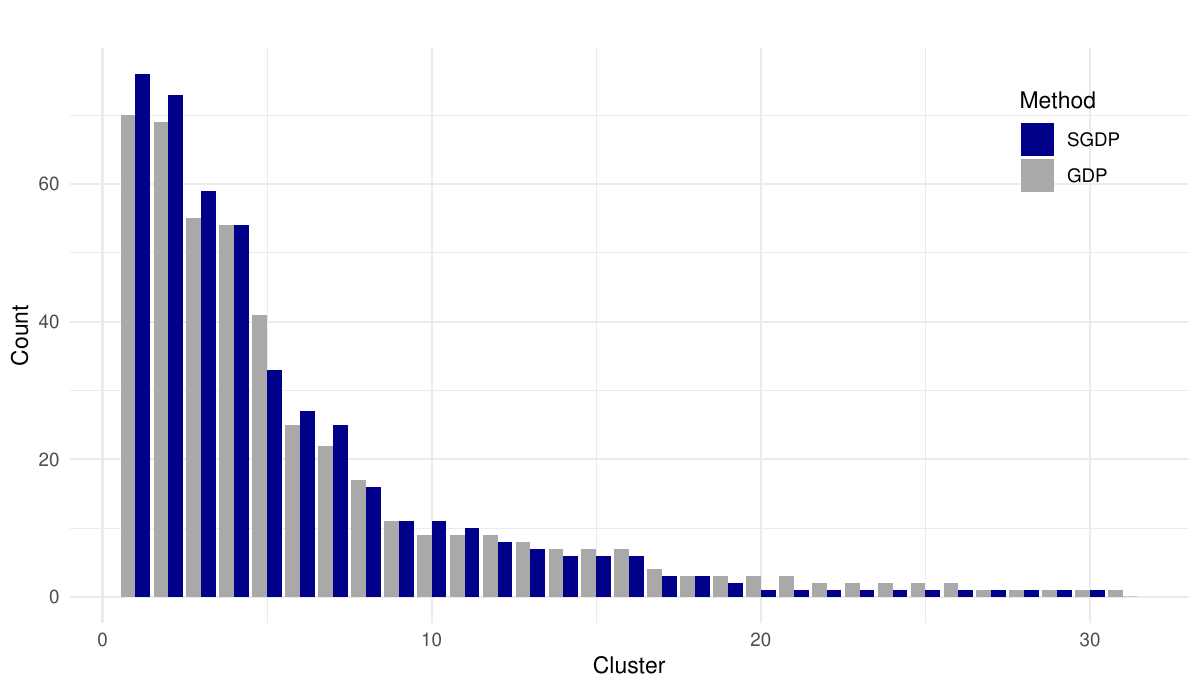}
    \caption{Distribution of the cluster sizes for pre-holiday effect obtained from the SGDP and GDP. }
    \label{fig:dist_ns3}
\end{figure}

\subsection{Spatial similarity}
Regarding the two clustering methods accounting for adjacencies, the resulting posterior means for $(\tau_1,\tau_2,\tau_3)$ were $(0.0097, 0.0176, 0.0125)$ for the SDP and $(0.0103, 0.0119, 0.0730)$ for the SGDP, indicating a notable correlation between geographically adjacent areas in the population data. To further examine this aspect, we compared the SGDP with the GDP. Figures~\ref{fig:dist_ns}--\ref{fig:dist_ns3} display the distributions of the SGDP and GDP. Both methods produced approximately the same number of clusters but differed in their proportions. As discussed in Section~\ref{subsec:prop}, this result indicates that the SGDP and GDP have identical probabilities of creating new clusters; however, the allocations within existing clusters differ. The detailed clusters are shown in Figure~\ref{tab:ns_plot}, where the three rows correspond to the three types illustrated in Figure~\ref{tab:cluster_plot}. The mean functions and clusters visible in the plots of the GDP are similar to those observed in the plots of the SGDP; however, they are relatively more dispersed than those of the SGDP. Notably, for the office and downtown areas, the clusters formed by the GDP overlook adjacent meshes owing to the absence of adjacency considerations. These findings underscore the importance of integrating spatial information into the methods.

\section{Discussion}\label{sec:dis}
This paper introduces a nonparametric Bayesian clustering method that infuses pairwise similarity into the GDP framework, effectively addressing high-dimensionality and spatial correlations. The method mitigates excess clusters resulting from both DP characteristics and high dimensionality by setting a prior distribution of the GDP parameters. The correlation of adjacent data is reflected in the similarity, the strength of which can be determined through the posterior distribution.

Additionally, the method encompasses temporal structures, demonstrating the ability to accurately track population clusters. The organization of information in clustering can be extended to other contexts. In particular, the clusters and mean functions identified by our method can serve as factors in factor models for predicting population data, as exemplified in \cite{wakayama2023spatiotemporal}. Future studies stand to benefit from methodologies that address temporal structure clustering for advanced practical applications, as seen in \cite{nieto2014bayesian,page2022dependent,deiorio2023bayesian}.

\section*{Acknowledgments}
This research was supported by JSPS KAKENHI (grant numbers 22J21090, 21H00699, 21K01421, 20H00080) and JST ACT-X (grant number JPMJAX23CS).

\bibliographystyle{chicago}
\bibliography{ref}

\begin{thebibliography}{}

\bibitem[\protect\citeauthoryear{Ahmadi-Javid, Berman, and Hoseinpour}{Ahmadi-Javid et~al.}{2018}]{ahmadi2018location}
Ahmadi-Javid, A., O.~Berman, and P.~Hoseinpour (2018).
\newblock {Location and Capacity Planning of Facilities with General Service-Time Distributions Using Conic Optimization}.
\newblock {\em arXiv preprint arXiv:1809.00080\/}.

\bibitem[\protect\citeauthoryear{Antoniak}{Antoniak}{1974}]{antoniak1974mixtures}
Antoniak, C.~E. (1974).
\newblock {Mixtures of Dirichlet Processes with Applications to Bayesian Nonparametric Problems}.
\newblock {\em The Annals of Statistics\/}~{\em 2\/}(6), 1152 -- 1174.

\bibitem[\protect\citeauthoryear{Ayed, Lee, and Caron}{Ayed et~al.}{2019}]{ayed2019beyond}
Ayed, F., J.~Lee, and F.~Caron (2019, 09--15 Jun).
\newblock {Beyond the Chinese Restaurant and Pitman-Yor processes: Statistical Models with double power-law behavior}.
\newblock In {\em Proceedings of the 36th International Conference on Machine Learning}, Volume~97, pp.\  395--404. PMLR.

\bibitem[\protect\citeauthoryear{Barcella, De~Iorio, Favaro, and Rosner}{Barcella et~al.}{2017}]{barcella2018dependent}
Barcella, W., M.~De~Iorio, S.~Favaro, and G.~L. Rosner (2017, 09).
\newblock {Dependent generalized Dirichlet process priors for the analysis of acute lymphoblastic leukemia}.
\newblock {\em Biostatistics\/}~{\em 19\/}(3), 342--358.

\bibitem[\protect\citeauthoryear{Connor and Mosimann}{Connor and Mosimann}{1969}]{connor1969concepts}
Connor, R.~J. and J.~E. Mosimann (1969).
\newblock {Concepts of independence for proportions with a generalization of the Dirichlet distribution}.
\newblock {\em Journal of the American Statistical Association\/}~{\em 64\/}(325), 194--206.

\bibitem[\protect\citeauthoryear{Cremaschi, Cadonna, Guglielmi, and Quintana}{Cremaschi et~al.}{2023}]{cremaschi2023change}
Cremaschi, A., A.~Cadonna, A.~Guglielmi, and F.~Quintana (2023).
\newblock A change-point random partition model for large spatio-temporal datasets.
\newblock {\em arXiv preprint arXiv:2312.12396\/}.

\bibitem[\protect\citeauthoryear{Dahl, Day, and Tsai}{Dahl et~al.}{2017}]{dahl2017random}
Dahl, D.~B., R.~Day, and J.~W. Tsai (2017).
\newblock {Random Partition Distribution Indexed by Pairwise Information}.
\newblock {\em Journal of the American Statistical Association\/}~{\em 112\/}(518), 721--732.

\bibitem[\protect\citeauthoryear{De~Blasi, Favaro, Lijoi, Mena, Pr{\"u}nster, and Ruggiero}{De~Blasi et~al.}{2013}]{deblasi2013gibbs}
De~Blasi, P., S.~Favaro, A.~Lijoi, R.~H. Mena, I.~Pr{\"u}nster, and M.~Ruggiero (2013).
\newblock {Are Gibbs-type priors the most natural generalization of the Dirichlet process?}
\newblock {\em IEEE Transactions on Pattern Analysis and Machine Intelligence\/}~{\em 37\/}(2), 212--229.

\bibitem[\protect\citeauthoryear{De~Iorio, Favaro, Guglielmi, and Ye}{De~Iorio et~al.}{2023}]{deiorio2023bayesian}
De~Iorio, M., S.~Favaro, A.~Guglielmi, and L.~Ye (2023).
\newblock Bayesian nonparametric mixture modeling for temporal dynamics of gender stereotypes.
\newblock {\em The Annals of Applied Statistics\/}~{\em 17\/}(3), 2256--2278.

\bibitem[\protect\citeauthoryear{Delaigle and Hall}{Delaigle and Hall}{2012}]{delaigle2012achieving}
Delaigle, A. and P.~Hall (2012).
\newblock Achieving near perfect classification for functional data.
\newblock {\em Journal of the Royal Statistical Society: Series B (Statistical Methodology)\/}~{\em 74\/}(2), 267--286.

\bibitem[\protect\citeauthoryear{Dunson and Johndrow}{Dunson and Johndrow}{2020}]{dunson2020hastings}
Dunson, D.~B. and J.~E. Johndrow (2020).
\newblock {The Hastings algorithm at fifty}.
\newblock {\em Biometrika\/}~{\em 107\/}(1), 1--23.

\bibitem[\protect\citeauthoryear{Ewens}{Ewens}{1972}]{ewens1972sampling}
Ewens, W.~J. (1972).
\newblock The sampling theory of selectively neutral alleles.
\newblock {\em Theoretical Population Biology\/}~{\em 3\/}(1), 87--112.

\bibitem[\protect\citeauthoryear{Ferguson}{Ferguson}{1973}]{ferguson1973bayesian}
Ferguson, T.~S. (1973).
\newblock {A Bayesian Analysis of Some Nonparametric Problems}.
\newblock {\em The Annals of Statistics\/}~{\em 1\/}(2), 209 -- 230.

\bibitem[\protect\citeauthoryear{Ferguson}{Ferguson}{1974}]{ferguson1974prior}
Ferguson, T.~S. (1974).
\newblock {Prior distributions on spaces of probability measures}.
\newblock {\em The Annals of Statistics\/}~{\em 2\/}(4), 615--629.

\bibitem[\protect\citeauthoryear{Folland}{Folland}{1999}]{folland1999real}
Folland, G.~B. (1999).
\newblock {\em {Real Analysis: Modern Techniques and Their Applications}\/} (2 ed.).
\newblock New York: John Wiley \& Sons.

\bibitem[\protect\citeauthoryear{Gelfand and Smith}{Gelfand and Smith}{1990}]{gelfand1990sampling}
Gelfand, A.~E. and A.~F. Smith (1990).
\newblock {Sampling-Based Approaches to Calculating Marginal Densities}.
\newblock {\em Journal of the American Statistical Association\/}~{\em 85\/}(410), 398--409.

\bibitem[\protect\citeauthoryear{Gelman, Carlin, Stern, Dunson, Vehtari, and Rubin}{Gelman et~al.}{2013}]{gelman2013bayesian}
Gelman, A., J.~B. Carlin, H.~S. Stern, D.~B. Dunson, A.~Vehtari, and D.~B. Rubin (2013).
\newblock {\em Bayesian Data Analysis\/} (3 ed.).
\newblock Chapman and Hall/CRC.

\bibitem[\protect\citeauthoryear{Glynn, Byrne, and Culhane}{Glynn et~al.}{2021}]{glynn2021inflection}
Glynn, C., T.~H. Byrne, and D.~P. Culhane (2021).
\newblock Inflection points in community-level homeless rates.
\newblock {\em The Annals of Applied Statistics\/}~{\em 15\/}(2), 1037--1053.

\bibitem[\protect\citeauthoryear{Grazian}{Grazian}{2023}]{grazian2023review}
Grazian, C. (2023).
\newblock {A review on Bayesian model-based clustering}.
\newblock {\em arXiv preprint arXiv:2303.17182\/}.

\bibitem[\protect\citeauthoryear{Hjort}{Hjort}{2000}]{hjort2000bayesian}
Hjort, N.~L. (2000).
\newblock {Bayesian analysis for a generalised Dirichlet process prior}.
\newblock {\em Preprint series. Statistical Research Report\/}.

\bibitem[\protect\citeauthoryear{Hubert and Arabie}{Hubert and Arabie}{1985}]{hubert1985comparing}
Hubert, L. and P.~Arabie (1985).
\newblock Comparing partitions.
\newblock {\em {Journal of Classification}\/}~{\em 2}, 193--218.

\bibitem[\protect\citeauthoryear{Ishwaran and James}{Ishwaran and James}{2001}]{ishwaran2001gibbs}
Ishwaran, H. and L.~F. James (2001).
\newblock {Gibbs Sampling Methods for Stick-Breaking Priors}.
\newblock {\em {Journal of the American Statistical Association}\/}~{\em 96\/}(453), 161--173.

\bibitem[\protect\citeauthoryear{Ishwaran and Zarepour}{Ishwaran and Zarepour}{2000}]{ishwaran2000markov}
Ishwaran, H. and M.~Zarepour (2000, 06).
\newblock {Markov chain Monte Carlo in approximate Dirichlet and beta two-parameter process hierarchical models}.
\newblock {\em Biometrika\/}~{\em 87\/}(2), 371--390.

\bibitem[\protect\citeauthoryear{Korwar and Hollander}{Korwar and Hollander}{1973}]{korwar1972contributions}
Korwar, R.~M. and M.~Hollander (1973).
\newblock {Contributions to the Theory of Dirichlet Processes}.
\newblock {\em The Annals of Probability\/}~{\em 1\/}(4), 705 -- 711.

\bibitem[\protect\citeauthoryear{Lu and Reddy}{Lu and Reddy}{2012}]{lu2012strategic}
Lu, A. and A.~Reddy (2012).
\newblock {Strategic Look at Friday Exceptions in Weekday Schedules for Urban Transit: Improving Service, Capturing Leisure Markets, and Achieving Cost Savings by Mining Data on Automated Fare Collection Ridership}.
\newblock {\em Transportation research record\/}~{\em 2274\/}(1), 30--51.

\bibitem[\protect\citeauthoryear{Lym}{Lym}{2021}]{lym2021exploring}
Lym, Y. (2021).
\newblock {Exploring dynamic process of regional shrinkage in Ohio: A Bayesian perspective on population shifts at small-area levels}.
\newblock {\em Cities\/}~{\em 115}, 103228.

\bibitem[\protect\citeauthoryear{MacEachern and M{\"u}ller}{MacEachern and M{\"u}ller}{1998}]{maceachern1998estimating}
MacEachern, S.~N. and P.~M{\"u}ller (1998).
\newblock {Estimating mixture of Dirichlet process models}.
\newblock {\em Journal of Computational and Graphical Statistics\/}~{\em 7\/}(2), 223--238.

\bibitem[\protect\citeauthoryear{Manning, Sch{\"u}tze, and Raghavan}{Manning et~al.}{2009}]{manning2009introduction}
Manning, C.~D., H.~Sch{\"u}tze, and P.~Raghavan (2009).
\newblock {\em {Introduction to Information Retrieval}}.
\newblock Cambridge University Press.

\bibitem[\protect\citeauthoryear{Miller and Harrison}{Miller and Harrison}{2013}]{miller2013simple}
Miller, J.~W. and M.~T. Harrison (2013).
\newblock {A simple example of Dirichlet process mixture inconsistency for the number of components}.
\newblock In {\em Advances in Neural Information Processing Systems}, Volume~26. Curran Associates, Inc.

\bibitem[\protect\citeauthoryear{Miller and Harrison}{Miller and Harrison}{2014}]{miller2014inconsistency}
Miller, J.~W. and M.~T. Harrison (2014).
\newblock {Inconsistency of Pitman-Yor Process Mixtures for the Number of Components}.
\newblock {\em Journal of Machine Learning Research\/}~{\em 15\/}(96), 3333--3370.

\bibitem[\protect\citeauthoryear{Mozdzen, Cremaschi, Cadonna, Guglielmi, and Kastner}{Mozdzen et~al.}{2022}]{mozdzen2022bayesian}
Mozdzen, A., A.~Cremaschi, A.~Cadonna, A.~Guglielmi, and G.~Kastner (2022).
\newblock {Bayesian modeling and clustering for spatio-temporal areal data: An application to Italian unemployment}.
\newblock {\em Spatial Statistics\/}~{\em 52}, 100715.

\bibitem[\protect\citeauthoryear{M{\"u}ller, Quintana, Jara, and Hanson}{M{\"u}ller et~al.}{2015}]{Muller2015}
M{\"u}ller, P., F.~A. Quintana, A.~Jara, and T.~Hanson (2015).
\newblock {\em Bayesian Nonparametric Data Analysis}.
\newblock Springer International Publishing.

\bibitem[\protect\citeauthoryear{Nagata, Aoyagi, and Kawakami}{Nagata et~al.}{2013}]{nagata2013using}
Nagata, T., S.~Aoyagi, and H.~Kawakami (2013).
\newblock Using mobile spatial statistics for regional revitalization.
\newblock {\em NTT DOCOMO Technical Journal\/}~{\em 14\/}(3), 46--50.

\bibitem[\protect\citeauthoryear{Neal}{Neal}{2000}]{neal2000markov}
Neal, R.~M. (2000).
\newblock {Markov Chain Sampling Methods for Dirichlet Process Mixture Models}.
\newblock {\em {Journal of Computational and Graphical Statistics}\/}~{\em 9\/}(2), 249--265.

\bibitem[\protect\citeauthoryear{Nieto-Barajas and Contreras-Crist{\'a}n}{Nieto-Barajas and Contreras-Crist{\'a}n}{2014}]{nieto2014bayesian}
Nieto-Barajas, L.~E. and A.~Contreras-Crist{\'a}n (2014).
\newblock {A Bayesian Nonparametric Approach for Time Series Clustering}.
\newblock {\em Bayesian Analysis\/}~{\em 9\/}(1), 147--170.

\bibitem[\protect\citeauthoryear{Oyabu, Terada, Yamaguchi, Iwasawa, Hagiwara, and Koizumi}{Oyabu et~al.}{2013}]{oyabu2013evaluating}
Oyabu, Y., M.~Terada, T.~Yamaguchi, S.~Iwasawa, J.~Hagiwara, and D.~Koizumi (2013).
\newblock Evaluating reliability of mobile spatial statistics.
\newblock {\em NTT DOCOMO Technical Journal\/}~{\em 14\/}(3), 16--23.

\bibitem[\protect\citeauthoryear{P{\'a}ez and Scott}{P{\'a}ez and Scott}{2004}]{paez2004spatial}
P{\'a}ez, A. and D.~M. Scott (2004).
\newblock {Spatial statistics for urban analysis: A review of techniques with examples}.
\newblock {\em GeoJournal\/}~{\em 61}, 53--67.

\bibitem[\protect\citeauthoryear{Page, Quintana, and Dahl}{Page et~al.}{2022}]{page2022dependent}
Page, G.~L., F.~A. Quintana, and D.~B. Dahl (2022).
\newblock {Dependent Modeling of Temporal Sequences of Random Partitions}.
\newblock {\em Journal of Computational and Graphical Statistics\/}~{\em 31\/}(2), 614--627.

\bibitem[\protect\citeauthoryear{Pitman}{Pitman}{1995}]{pitman1995exchangeable}
Pitman, J. (1995).
\newblock Exchangeable and partially exchangeable random partitions.
\newblock {\em Probability Theory and Related Fields\/}~{\em 102\/}(2), 145--158.

\bibitem[\protect\citeauthoryear{Pitman}{Pitman}{1996}]{pitman1996some}
Pitman, J. (1996).
\newblock {Some developments of the Blackwell-MacQueen urn scheme}.
\newblock {\em Lecture Notes-Monograph Series\/}, 245--267.

\bibitem[\protect\citeauthoryear{Pitman and Yor}{Pitman and Yor}{1997}]{pitman1997two}
Pitman, J. and M.~Yor (1997).
\newblock {The two-parameter Poisson-Dirichlet distribution derived from a stable subordinator}.
\newblock {\em The Annals of Probability\/}~{\em 25\/}(2), 855 -- 900.

\bibitem[\protect\citeauthoryear{Pol}{Pol}{1986}]{pol1986marketing}
Pol, L.~G. (1986).
\newblock Marketing and the demographic perspective.
\newblock {\em Journal of Consumer Marketing\/}~{\em 3\/}(1), 57--65.

\bibitem[\protect\citeauthoryear{Rasmussen and Williams}{Rasmussen and Williams}{2006}]{rasmussen2006gaussian}
Rasmussen, C.~E. and C.~K.~I. Williams (2006).
\newblock {\em {Gaussian Processes for Machine Learning}}.
\newblock MIT press Cambridge, MA.

\bibitem[\protect\citeauthoryear{Rodriguez and Dunson}{Rodriguez and Dunson}{2014}]{rodriguez2014functional}
Rodriguez, A. and D.~B. Dunson (2014).
\newblock {Functional clustering in nested designs: Modeling variability in reproductive epidemiology studies}.
\newblock {\em The Annals of Applied Statistics\/}~{\em 8\/}(3), 1416 -- 1442.

\bibitem[\protect\citeauthoryear{Sethuraman}{Sethuraman}{1994}]{sethuraman1994constructive}
Sethuraman, J. (1994).
\newblock {A constructive definition of Dirichlet priors}.
\newblock {\em Statistica Sinica\/}~{\em 4\/}(2), 639--650.

\bibitem[\protect\citeauthoryear{Stutz}{Stutz}{2004}]{stutz2004charting}
Stutz, F.~P. (2004).
\newblock Charting urban travelers 24--7 for disaster evacuation and homeland security.
\newblock In {\em WorldMinds: Geographical Perspectives on 100 Problems: Commemorating the 100th Anniversary of the Association of American Geographers 1904--2004}, pp.\  177--182. Springer.

\bibitem[\protect\citeauthoryear{Wade and Ghahramani}{Wade and Ghahramani}{2018}]{wade2018bayesian}
Wade, S. and Z.~Ghahramani (2018).
\newblock {Bayesian Cluster Analysis: Point Estimation and Credible Balls (with Discussion)}.
\newblock {\em Bayesian Analysis\/}~{\em 13\/}(2), 559--626.

\bibitem[\protect\citeauthoryear{Wakayama and Imaizumi}{Wakayama and Imaizumi}{2024}]{wakayama2021fast}
Wakayama, T. and M.~Imaizumi (2024).
\newblock {Fast Convergence on Perfect Classification for Functional Data}.
\newblock {\em Statistica Sinica\/}~{\em 34\/}(4).

\bibitem[\protect\citeauthoryear{Wakayama and Sugasawa}{Wakayama and Sugasawa}{2024}]{wakayama2023spatiotemporal}
Wakayama, T. and S.~Sugasawa (2024).
\newblock Spatiotemporal factor models for functional data with application to population map forecast.
\newblock {\em Spatial Statistics\/}, 100849.

\bibitem[\protect\citeauthoryear{Wang and Mu}{Wang and Mu}{2018}]{wang2018spatial}
Wang, M. and L.~Mu (2018).
\newblock {Spatial disparities of Uber accessibility: An exploratory analysis in Atlanta, USA}.
\newblock {\em Computers, Environment and Urban Systems\/}~{\em 67}, 169--175.

\bibitem[\protect\citeauthoryear{Xie, Yu, Zheng, Wang, and Jiang}{Xie et~al.}{2021}]{xie2021revealing}
Xie, C., D.~Yu, X.~Zheng, Z.~Wang, and Z.~Jiang (2021).
\newblock {Revealing spatiotemporal travel demand and community structure characteristics with taxi trip data: A case study of New York City}.
\newblock {\em PLoS one\/}~{\em 16\/}(11), e0259694.

\bibitem[\protect\citeauthoryear{Zhang, Fu, Kong, and Zhang}{Zhang et~al.}{2019}]{zhang2019prefecture}
Zhang, Y., Y.~Fu, X.~Kong, and F.~Zhang (2019).
\newblock {Prefecture-level city shrinkage on the regional dimension in China: Spatiotemporal change and internal relations}.
\newblock {\em Sustainable Cities and Society\/}~{\em 47}, 101490.

\end{thebibliography}

\end{document}